\documentclass[reprint,twocolumn,prd,aps,amsmath,amssymb,floatfix,nofootinbib,superscriptaddress]{revtex4-2}

\usepackage{graphicx}
\usepackage{dcolumn}
\usepackage{bm}
\usepackage[usenames, dvipsnames]{color}
\usepackage[normalem]{ulem}
\usepackage[dvipsnames]{xcolor}
\usepackage[utf8]{inputenc}
\usepackage{hyperref}
\usepackage{aas_macros}
\usepackage{multirow}
\usepackage{comment}
\usepackage{cancel}
\usepackage{bm}
\usepackage{subfig}
\usepackage{soul}

\usepackage{caption}
\usepackage[toc,page]{appendix}

\newcommand{\be}{\begin{align}}

\newcommand{\ee}{\end{align}}
\newcommand*\dd{\mathop{}\!\mathrm{d}}
\newcommand{\msu}{eXtreme Gravity Institute, Department of Physics,\\
Montana State University, Bozeman, Montana 59717, USA}
\newcommand{\gw}{gravitational-wave }
\newcommand{\gws}{gravitational-waves }
\newcommand{\Gw}{Gravitational-wave }

\begin{document}

\title{Handling Data Gaps for the Next Generation of Gravitational-Wave Observatories}

\author{Noah~Pearson}
\affiliation{\msu}

\author{Neil J. Cornish}
\affiliation{\msu}


\begin{abstract}

    In the coming decades, as the low frequency sensitivity of detectors improves, the time that \gw signals remain in the sensitive band will increase, leading to new challenges in analyzing data, namely non-stationary noise and data gaps. Time-frequency (wavelet) methods can efficiently handle non-stationary noise, but data gaps still lead to spectral leakage due to the finite length of the wavelet filters. It was previously shown that Bayesian data augmentation - ``gap filling" - could mitigate spectral leakage in frequency domain analyses, but the computational cost associated with the matrix operations needed in that approach is prohibitive. Here we present a new, computationally efficient approach to Bayesian data augmentation in the time-frequency domain that avoids repeated, costly matrix operations. We show that our approach efficiently solves the problem of data gaps in simulated LISA data, and can be smoothly integrated into the LISA Global Fit. The same approach can also be used for future 3G ground-based interferometers.
    
\end{abstract}

\maketitle

\section{Introduction}


The future decades of \gw astronomy offer much promise for discovery; we must be prepared for the challenges in analyzing data that certainly await us. Most immanently, the space-borne Laser Interferometer Space Antenna (LISA)~\cite{LISA:2024hlh} is scheduled to launch in the 2030's. It is composed of a triad of co-moving satellites in a large ($2.5 \times 10^6$ km) triangular constellation that will trail the Earth in its orbit, designed to detect \gws in the frequency band ($10^{-4}-1$ Hz) intermediate between the LIGO and Pulsar Timing Array bands. LISA presents a novel window on the \gw spectrum and will grant access to new sources, and new insights into known ones. In particular, LISA will be sensitive to a multitude of compact binaries ($\mathcal{O}\left[10^4\right]$) within the Milky Way known generally as ``galactic binaries." This vast population of sources accessible to LISA offers an unprecedented opportunity for a comprehensive survey of our galaxy~\cite{lisa_prop,lisa_white,lisa22}. Beyond observing our home galaxy, LISA will be able to study extragalactic signals from supermassive black hole binaries, and primordial sources from early in the Universe's history~\cite{Ricciardone2016,Cornish2006}. To realize LISA's full science potential, we address the problem of data gaps and disturbances.

 LISA and other planned observatories will present a number of analysis challenges, among them non-stationary noise and gaps in the data time series~\cite{Baghi_2019,Digman2022}. The standard Fourier domain analyses will become computationally intractable due to these features given the large datasets produced by these detectors. The future of \gw detection drives us to longer-lived signals. In ground-based detectors, improved low-frequency sensitivity means signals will be observable earlier in their evolution, and space-based detectors target sources that evolve slowly in frequency and will be in-band for many months or years~\cite{lisa_prop,lisa_white}. The long duration of signals exposes modern analyses to two significant challenges: noise spectra can no longer be approximated as stationary, and gaps in the data will coincide with signals. Non-stationary noise renders traditional Fourier domain analyses computationally intractable, since the noise correlation matrix becomes dense, leading to a prohibitive computational cost. Data gaps present a challenge of a different nature; as discontinuities in the data, they can introduce spectral leakage when working outside of the time domain. If not properly accounted for, this can bias the analysis. Previously, these gaps were not as concerning for ground-based detectors since gaps were short and infrequent enough that they were not likely to coincide with a signal. In LISA and future detectors, this is not the case. These gaps can lead to biases in the analyses, that can be further compounded by potential mis-modeling of the noise. The gaps also pose a significant computational challenge. Therefore, new tools must be developed to ameliorate these obstacles for the next generation of \gw observatories.



There have been a number of studies on managing data gaps, for both ground-based and space-based detectors. In evaluating which technique will be most effective, a balance must be found between a method's effectiveness at suppressing spectral leakage and computational cost. Among the solutions so far explored are apodization techniques~\cite{dey2021,Burke2025a,Castelli2024}, ``gating and inpainting" methods (primarily in the context of ground-based detectors)~\cite{Venumadhav2019,zackay2021,Capano:2021,Isi2021,Blelly_2021,wang24}, imputation via deep learning models~\cite{Mao2024}, and Bayesian data augmentation~\cite{Baghi_2019}. All treatments thus far explored have been applied to stationary noise processes in the Fourier domain or in the time domain~\cite{Burke2025a}. 

Among the methods that are most effective at removing spectral leakage, Bayesian data augmentation stands out, albeit with greater computational cost. The method works by imputing (simulating) data to fill the gaps. In Bayesian data augmentation the gaps are filled with simulated data that is drawn from a distribution conditioned on the observed data. This method accounts for correlations from the observed noise, and optimally mitigates spectral leakage. A nice feature of this approach is that the observed data is left untouched (preserving the observed SNR), and for stationary noise, the noise covariance retains its circulant structure (allowing for Whittle's approximation to the likelihood), so no other aspects of the analysis need be modified (e.g. the form of signal templates), and it may be implemented in a Bayesian setting without bias whereby the imputing data is updated as the noise and signal models are explored. The drawback to this approach is primarily the computational complexity, since the evaluation of the conditional distribution requires costly matrix operations~\cite{Baghi_2019}.


Herein, we outline a computationally efficient approach to Bayesian data augmentation applied to time-frequency (wavelet) domain analyses. Working in the wavelet domain allows us to efficiently handle a broad swath of non-stationary noise processes, and our application of data augmentation makes gap filling an efficient, accurate, and feasible solution. We also show its usefulness in eliminating spectral leakage due to the edges in a finite time series, and its application even in the presence of a noise state change across a gap. Beyond this work's intended use in LISA, we expect it to be general enough to be applicable in any future experiment that encounters the same issues of non-stationarity and data gaps.

We introduce several innovations to make Bayesian data augmentation computationally efficient:
\begin{itemize}
  \item Costly matrix operations are avoided by sampling the conditional distribution for the imputed data via a Markov Chain Monte Carlo rather than directly computing the conditional distribution.
  \item Performing the data augmentation in the wavelet domain rather than the Fourier domain, which isolates the effects of the gaps and allows for non-stationarity in the noise.
  \item Utilize specialized fast wavelet transform techniques to
  efficiently map between the time domain and wavelet domain.
\end{itemize}


The paper is structured as follows:
in Sec.(\ref{sec: data gaps}) we review the likely causes and effects gaps will have on analyses left unaddressed, in Sec.(\ref{sec: data aug}) we describe our process of augmenting data, in Sec.(\ref{sec: tf analysis}) we motivate the need to perform Bayesian inference in the time-frequency domain, and in Sec.(\ref{sec:simulation},\ref{sec: results}) we demonstrate and discuss this new prototype tool as applied to parameter estimation in a LISA-like simulation, and show its other useful extensions in analyzing \gw data. We also provide appendices that deal in finer detail with the implementation of our procedure.

\section{Data Gaps}
\label{sec: data gaps}

\subsection{Cause for Pause}

The reasons for gaps in the data are entirely irrelevant for the application of this work, they simply inform our expectations and simulations. Our tool may be applied to any quantity or size of gaps. To further emphasize the severity of the problem of gaps and the necessity of this study, and to motivate the design of the simulated data we will test on, we will briefly review the causes of data gaps in the most pressing experiment. Since we intend this tool for future detectors, we cannot speak specifically on the cause for or demographics of data gaps except in LISA where we have some prior experiments and matured designs in place; especially now that it has passed the adoption milestone with the European Space Agency (ESA)~\cite{LISA2024}. 


LISA data will undergo several levels of processing prior to a global fit analysis to extract astrophysical signals. The data that we will receive will be in the form of Time Delay Interferometry (TDI) variables - combinations of measurements of the laser phase constructed primarily to suppress laser frequency noise - which constitute a time series~\cite{LISA2024}. Like any other instrument, LISA will not be perfect, and will likely experience noise artifacts, glitches, and gaps produced either by onboard electronics, by interactions with its environment (e.g. micrometeoroid collisions), or instrument maintenance. These varieties of corruption in the data will be both predictable and random in their occurrence, and corrupt the data over extended and short time scales. It is conceivable that with the diversity of corruption, many analysis tools specialized to each variety could be used in tandem.


For noise artifacts or glitches, one approach is to perform glitch subtraction, as done for non-Gaussian features in \texttt{BayesWave}~\cite{Cornish2014}, and in other studies~\cite{zackay2021,Muratore2025}. Alternatively, the transient noise aberrations may make the data unusable, in which case one could choose to window out the corrupted segment of data - thereby creating a ``gap" in the time series. In fact, one could window out any variety of corruption to be conservative and introduce more gaps; these aren't \textit{true} gaps, rather voluntary ones. The LISA Pathfinder mission experienced glitches with nearly daily frequency~\cite{pathfinder}. Following the prescription of Baghi \textit{et al.}, 10 minutes of data surrounding these glitches would be excised to conservatively exclude their influence on neighboring data~\cite{Baghi_2019}. Other unpredictable causes of gaps will not constitute a large fraction of the data lost~\cite{Castelli2024,Burke2025b}. Unpredictable true gaps may occur due to instrument malfunctions and minor outages, lasting approximately 10 seconds~\cite{Burke2025a}. Random micrometeroid collisions are expected to contribute to less than 10\% of mission down time~\cite{LISA2024}, but each collision may lose about a day of data~\cite{Burke2025a}.


Another potential source of gaps is maintenance on the instruments. Throughout LISA's mission life, data collection will be interrupted to perform various routine measures to maintain the optimum sensitivity of the detector, such as antenna re-alignment (gap duration $\sim$ 3 hours every $\sim$ 2 weeks), tilt-to-length coupling constant estimation (gap duration $\sim$ 2 days every $\sim$ 3 months), and point-ahead angle mechanism adjustments (PAAM) (gap duration $\sim$ 1 second every $\sim$ 8 hours - 12-20 days, M. Hewitson \& G. Mueller pers. comm.), etc.~\cite{LISA2024,Burke2025a}. It's possible these operations don't amount to a total loss of data, rather only elevated or un-characterized noise; in either case, the data are likely unsuitable to be included in the analysis during these periods and should be omitted. In the case of PAAM, these disturbances could be treated as noise transients upon an otherwise stable noise background, that are modeled and subtracted as is done by the \texttt{BayesWave}~\cite{Cornish2014} algorithm.


An important distinction must be made, \textit{true} gaps that occur in the instrument prior to the formation of TDI variables will be enlarged by 100 seconds, and will have smoothly-tapered edges, rather than sharp edges (Eleonora Castelli pers. comm.)~\cite{Burke2025b, LISA2024}. In applying this tool to TDI variables (left to future work), one could choose to remove the edge smoothing effect and treat only gaps with sharp edges.  

In total, we can expect that gaps will cost a fair fraction of LISA's observation time (and SNR). The most recent estimates place the mission duty cycle at $> 82\%$~\cite{LISA2024}. In the following section, we will review the issues these gaps present in an analysis, and the strategies to manage them.

\subsection{Adverse Effects and Solutions}

A gap is a continuous segment of absent (or zeroed) data, with sharp transitions to observed data at the boundaries of the segment. They irreparably cost us information and SNR. Were we to perform analyses in the time domain, these gaps would present no other issue; one could simply avoid gap segments in the likelihood calculation. This approach is explored in the context of ``all-in-one" TDI-$\infty$~\cite{Houba2024}. Generally speaking, this would be computationally impractical, since the noise covariance matrix in the time domain would be dense and expensive to invert. Houba \textit{et al.} attempt to circumvent this issue by breaking the TDI measurement time series into chunks, thereby reducing the size of the covariance matrix, but at the hazard of neglecting correlations. This can be mitigated by designing the chunks to have some overlap. 

Traditionally, gravitational wave analyses have been performed in the Fourier domain, where the noise covariance matrix is diagonal for stationary noise. In the presence of gaps, this transformation introduces the well-understood pathology of ``spectral leakage", where power in the Fourier spectrum is spread to different frequency bins due to the sharp features of the gaps. This introduces additional correlations in the Fourier domain which again make the covariance matrix dense. Making the approximation that the covariance is diagonal would ignore these correlations and is in general not valid for gapped data~\cite{Baghi_2019}. Our challenge is then to address spectral leakage. There are many options available to handle this problem, that vary in computational complexity and effectiveness; we must find a balance between the two. 

The simplest treatment is apodization - the application of a smoothing function which reduces spectral leakage by rounding-off the edges of gaps. In general for infrequent gaps, apodization is not problematic. However, the application of a smoothing function would cost significant SNR\footnote{There is loss in SNR due to a window since this method does not also introduce a window into the noise model.} due to the long rise-time of the function needed in LISA data; this is especially true for frequent, short gaps. This technique has also been explored in conjunction with marginalizing the likelihood over missing data~\cite{Burke2025a}. Burke \textit{et al.} show that by incorporating the apodization into the noise covariance matrix, that the SNR of apodized data is the same as gapped data. This however requires a Moore-Penrose pseudo-inverse of the noise covariance to translate their method into the Fourier domain, and data segments limited in size since it still requires a dense noise covariance matrix.

Alternatively, leakage could be accounted for by computing gapped signal templates. Because they themselves leak, this methods obviates the benefit of having a fast signal transform (to be discussed). Additionally, one would have to compute gapped noise models as well, which would no longer be Gaussian. While this approach is possible, it is far from ideal.

To avoid leakage entirely, one could modify the functional basis. Spectral leakage may also be interpreted as the breaking of functional basis orthogonality around the edges of a gap, since a portion of a particular basis function may extend into the gapped region. The time-frequency extent of its effect is determined by the properties of the basis transform kernel. Instead, we could choose to construct a basis that maintains orthogonality around a gap; it would require a specialized set of basis functions unique to each gap. While complex, it also would compromise other desirable properties of the basis, in our case the time-frequency localization (see Sec.(\ref{sec: tf analysis})).

Among the solutions so far popularly explored are ``gating and inpainting" methods (primarily in the context of ground-based detectors)~\cite{Venumadhav2019,zackay2021,Capano:2021,Isi2021,Blelly_2021,wang24}. Data gating and inpainting works by applying an apodizing window to gaps, and filling in gaps with an interpolation that by requirement doesn't modify the likelihood. We claim this is not a desirable solution, particularly if data gaps (even of short duration) are frequent in the data time series, since much of the SNR will be lost to apodization. Additionally, windowing gapped data ruins the circulant property of the noise covariance matrix, thus precluding the use of Whittle's approximation to the likelihood~\cite{wang24}. The imputed interpolating data also is not required to be statistically consistent with the observed data, and does not fully eliminate spectral leakage, even if the noise Power Spectral Density (PSD) is known exactly. 

Machine learning has also been brought to bear on this issue. Differing from the prior inpainting techniques, deep learning models may be trained to generate infilling data. The most recent study has only performed on noiseless massive black hole inspiral signals, with the assumption that denoising processes have already taken place in the analysis~\cite{Mao2024}. When a gap is placed far from merger they recover accurate signal model parameters, with a minor bias when the gap occurs at merger.

Another method, most pertinent to this work, is known as Bayesian data augmentation. In this technique, data is imputed in gaps by drawing points from a distribution conditioned on the observed data~\cite{Baghi_2019}. This accurately builds in correlations from the observed noise, and optimally mitigates spectral leakage. What's further appealing about this method is that the observed data is left untouched (preserving SNR in the observed data versus methods involving apodization and diagonal covariances), Whittle's approximation is valid, and it may be implemented in a Bayesian setting whereby the imputing data is updated as the posteriors on the signal and noise are simultaneously explored. The drawback to this approach is primarily the computational complexity, since the evaluation of the conditional distribution with which to draw imputing data requires many costly matrix operations. And, common to all methods thus far explored, it has been done in the standard frequency domain which will be intractable for future detectors.

In the next sections, we introduce a modification to Bayesian augmentation that makes it far more practical, and motivate the choice to perform the analysis in the time-frequency domain.

\section{Data Augmentation}
\label{sec: data aug}

\subsection{Data Imputation}

Consider a complete data time series $\bm{d}$ of length $N$. With gaps in it, we only observe $N_o$ of those points, and are missing $N_m$ points (i.e. $N_o+N_m=N$). It is convenient to define two rectangular masking matrices which map the complete data to both the observed data $\bm{d}_o$ and missing data $\bm{d}_m$,

\begin{equation}
    \begin{split}
    & \bm{d}_o = \bm{M}_o \bm{d}\\  
    & \bm{d}_m = \bm{M}_m \bm{d}\,,
    \end{split}
\end{equation}

\noindent such that $\bm{d} = \bm{d}_o \oplus \bm{d}_m$, where we define $\oplus$ as the time-ordered concatenation operation. 

Data imputation is the insertion of artificial data $\bm{d}_i$ into the gaps of observed data, in place of missing data, thereby ``augmenting" it $\bm{d}_A$,

\begin{equation}
    \bm{d}_A = \bm{d}_o \oplus \bm{d}_i\,.
\end{equation}

It is important to note that augmenting data does \textit{not} enhance the SNR, as no new information is added; we show results to this effect in Sec.(\ref{sec: results}). Conceivably, the artificial ``imputing" data can be chosen arbitrarily. However, without proper care it can bias one's analysis or do little to mitigate the effects of gaps while working outside of the time-domain. We choose to follow the method of Baghi \textit{et al.}, who imputed statistically-consistent data as drawn from a multivariate Gaussian distribution conditioned on the observed data, $p\,(\bm{d}_m|\bm{d}_o,\bm{\theta})$~\cite{Baghi_2019}. This only assumes that the data is Gaussian - it may be non-stationary. By imputing such data, the likelihood will be unchanged, and the spectral leakage is optimally mitigated. This distribution has mean and covariance~\cite{Baghi_2019, eaton1983},

\begin{equation}
    \begin{cases}
        & \bm{\mu}_{m|o} = \bm{h}_m(\bm{\theta}) + \bm{\bm{\Sigma}}_{mo}\bm{\Sigma}_{oo}^{-1}(\bm{d}_o-\bm{h}_o(\bm{\theta}))\\
        & \bm{\Sigma}_{m|o} = \bm{\Sigma}_{mm} - \bm{\Sigma}_{mo}\bm{\Sigma}_{oo}^{-1}\bm{\Sigma}_{mo}^\mathrm{T}\,,
    \end{cases}
    \label{eq: conditional dist}
\end{equation}

\noindent where $\bm{\Sigma}_{mo} = \bm{M}_m\bm{\Sigma} \bm{M}_o^\mathrm{T}$, the mixed covariance between the missing and observed data, $\bm{\Sigma}_{mm} = \bm{M}_m\bm{\Sigma} \bm{M}_m^\mathrm{T}$, and $\bm{\Sigma}_{oo} = \bm{M}_o\bm{\Sigma} \bm{M}_o^\mathrm{T}$, the covariances of the missing and observed data, respectively ($\bm{\Sigma}$ is the complete time domain covariance). $\bm{h}_o(\bm{\theta}), \bm{h}_m(\bm{\theta})$ are the observed and missing segments of the signal model, respectively. The imputed data is generated in the time domain, and then returned to the analysis' domain of choice. Drawing data from a multivariate normal distribution requires the square-root of the covariance matrix, e.g. the Cholesky decomposition, $\bm{\Sigma}_{m|o} = \bm{HH}^\mathrm{T}$,

\begin{equation}
    \bm{d}_i = \bm{\mu}_{m|o} + \bm{H} \bm{\gamma}\,,
    \label{eq: impute data}
\end{equation}

\noindent where $\bm{\gamma}$ is a $N_m\times1$ vector of Gaussian random deviates with zero mean and unit variance (Gaussian white noise); correlations are imparted by the covariance matrix. We see from this form that there are two components to the imputing data: deterministic and stochastic. There exists a mean that continuously connects both ends of a gap, and a colored noise overlaid which imbues the mean with the proper correlations and vanishes at the gap boundaries. Another method is to use the Eigen-decomposition of the conditional covariance matrix. 

One may be concerned that the insertion of artificial data would bias an analysis. In general, that concern is appropriate. However, as applied in a Bayesian setting it's easy to demonstrate this is unfounded. The imputing data may be broken into two parts, contributions from the noise and signal,

\begin{equation}
    \bm{d}_i = \bm{n}_i + \bm{h}_i\,.
\end{equation}

\noindent In the likelihood function, the signal model is subtracted directly from the data,

\begin{equation}
    \mathrm{ln}\mathcal{L}(\bm{\theta}|\bm{d}_A) \propto -\frac{1}{2}(\bm{d}_A-\bm{h}(\bm{\theta}))^\dagger \bm{\Sigma}^{-1}(\bm{d}_A-\bm{h}(\bm{\theta}))\,.
\end{equation}

\noindent This means that the imputing signal $\bm{h}_i = \bm{h}_m(\bm{\theta})$ is removed from the augmented data; its only function was to eliminate spectral leakage, it does not contribute to the likelihood. The noise component of the augmented data is also subtracted via the noise covariance matrix, i.e. ``subtraction equals division."~\cite{Lentati2013,Cornish2013,Romano2016} Unfortunately, generating imputing data in this way is computationally costly, but we propose an efficient approach to integrating data augmentation into Bayesian analyses.

\subsection{Novel Sampling Algorithm}
\label{sec: novel sampling algo}



This augmentation method requires us to make draws from the conditional distribution, $p\,(\bm{d}_m|\bm{d}_o,\bm{\theta})$. The most costly step in this method is calculating the conditional distribution's mean and covariance (specifically, $\bm{\Sigma}_{oo}^{-1}$). As implemented in Baghi \textit{et al.}, who applied data augmentation in the context of Bayesian parametric inference, the covariance must properly be updated every advance of the Markov Chain Monte Carlo (MCMC) state (the authors however perform this update every 100 advances). In this way, data may be imputed with a direct draw - rather than having to evaluate the Metropolis-Hastings ratio - but it requires the repeated computation of the conditional distribution's moments. Updating the covariance matrix and the conditional distribution in general scales as the cube of the number of data points.\footnote{the authors chose to limit the covariance calculation to a ``neighborhood" around each gap, cutting down on cost. However, working in a Fourier basis which is non-local, this treacherously neglects large-scale (red) correlations present in the data. Later, we will make the same approximation. But, while in the previous work by Baghi \textit{et al.} this is effectively an approximation on the \textit{posterior}, here it is only an approximation on the \textit{proposal}.}

We propose a novel way of drawing imputing data which avoids these frequent costly matrix operations and still retains the authentic statistical character of the data. Imputing data may be drawn using an MCMC; the likelihood will be the arbiter in accepting or rejecting imputing data. In principle, imputing data can be generated with any proposal, but unless the proposals produce fills that are consistent with the surrounding data, they are likely to have very low acceptance rates. The most efficient proposal is the conditional distribution itself. 

We choose some previous state of the MCMC $\bm{\overline{\Sigma}}$ to calculate the distribution's mean and covariance and use it as the imputation proposal distribution. The imputation proposal distribution may be updated very infrequently, to reduce the cost of constantly recomputing it. Even if the noise model is incorrect, this is merely a proposal, so as long as it is broadly consistent with a draw from the posterior, it will have a high acceptance rate. And, as the MCMC evolves, it will ``burn-in" to the truth and improve its acceptance rate as the Metropolis-Hastings ratio approaches unity.

This method performs the traditional Metropolis-Hastings algorithm common in MCMC techniques, with the complication that the data that appears in the likelihood is subject to change throughout the posterior sampling, as a result of augmentation. We believe this abnormality benefits from further clarification. 

Because for each noise imputation, the filling data $\bm{x}_m$ are a completely new draw from the conditional Gaussian distribution with a non-zero mean (cf. Eq.(\ref{eq: conditional dist})), it does not have a symmetric proposal probability,

\begin{widetext}
\begin{eqnarray}
        p\,(\bm{x}_m |\bm{d}_o,\bm{\Sigma}) &=& \frac{1}{\sqrt{(2\pi)^{N_m}|\bm{\Sigma}_{m|o}}|}
        \nonumber
         \mathrm{exp}\left[-\frac{1}{2}(\bm{x}_m - \bm{\mu}_n)^\mathrm{T} \bm{\Sigma}_{m|o}^{-1} (\bm{x}_m-\bm{\mu}_n)\right]\,,
\end{eqnarray}
\end{widetext}

\noindent where $\bm{x}_m$ is the missing, imputed noise vector drawn from the conditional distribution, and $\bm{\mu}_n = \bm{\Sigma}_{mo}\bm{\Sigma}_{oo}^{-1}\bm{d}_o$, the noise portion of the mean of the imputed data. Because the probability is not symmetric between two fills $\bm{x}_m, \bm{y}_m$, this means that it doesn't cancel in the Metropolis-Hastings ratio. It is fortunately a cheap quantity to compute, as the inverse of the conditional noise covariance is only $\mathcal{O}(N_m^3)$, where $N_m$ is the number of missing data points in the gap being updated.

It is easy to show that in the case that the noise model is unchanged and is the same for the fill proposals as well as the likelihoods, that any filling draw from the conditional distribution is valid, as in the work of Baghi \textit{et al.}~\cite{Baghi_2019}. Moving from imputation $\bm{x}_m\rightarrow \bm{y}_m$, the Metropolis-Hastings ratio is (assuming uniform priors, which cancel),

\begin{equation}
    R = \frac{p\,(\bm{y}_m,\bm{d}_o|\bm{\Sigma})\,p\,(\bm{x}_m|\bm{d}_o,\overline{\bm{\Sigma}})}
    {p\,(\bm{x}_m,\bm{d}_o|\bm{\Sigma})\,p\,(\bm{y}_m|\bm{d}_o,\overline{\bm{\Sigma}})}\,,
\end{equation}

\noindent where $p\,(\bm{x}_m,\bm{d}_o|\bm{\Sigma})$ are the likelihoods, and $p\,(\bm{x}_m |\bm{d}_o,\overline{\bm{\Sigma}})$ are the asymmetric fill proposal probabilities. Note that in general $\bm{\Sigma} \neq \overline{\bm{\Sigma}}$. Knowing that $p\,(\bm{x}_m,\bm{d}_o|\bm{\Sigma}) = p\,(\bm{d}_o)\,p\,(\bm{x}_m|\bm{d}_o,\bm{\Sigma})$ (and similarly for $\bm{y}_m$), the ratio reduces to,

\begin{equation}
    R = \frac{p\,(\bm{y}_m|\bm{d}_o,\bm{\Sigma})\,p\,(\bm{x}_m|\bm{d}_o,\overline{\bm{\Sigma}})}
    {p\,(\bm{x}_m|\bm{d}_o,\bm{\Sigma})\,p\,(\bm{y}_m|\bm{d}_o,\overline{\bm{\Sigma}})}.
\end{equation}

\noindent In the work of Baghi \textit{et al.}, the conditional covariance is (properly) re-computed (at great cost) every advance of the MCMC state; so $\bm{\Sigma} = \overline{\bm{\Sigma}}$, thus $R = 1$. In that case, any noise fill would be accepted with 100\% probability in the Metropolis-Hastings algorithm. However, in our implementation, the covariances used in the likelihoods and fill proposals will be different, and $R \neq 1$ in general. However, conceivably for appropriate proposals in the noise model that are broadly consistent with the posterior distribution, the likelihood ratio won't be all that different from that of the proposal probabilities, and $R \sim 1$, leading to good acceptance that will only improve as $\overline{\bm{\Sigma}}\rightarrow\bm{\Sigma}$ over some burn-in period. 

\begin{figure}
    \centering
    \includegraphics[width=1\linewidth]{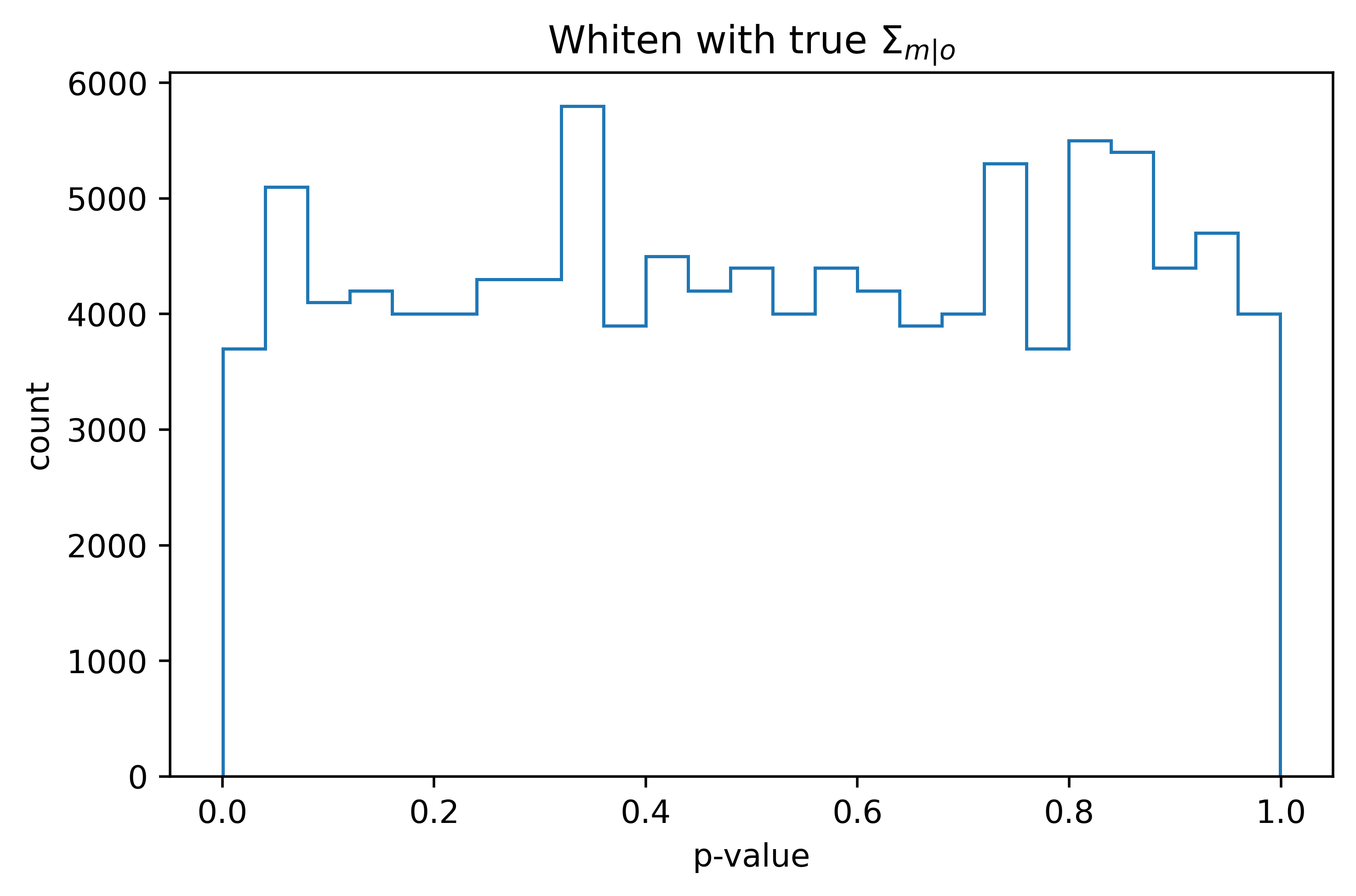}
    \caption{Distribution of p-values for the Anderson-Darling test of normality on fill data generated using our sampling scheme, whitened with the true conditional noise covariance. The uniform distribution demonstrates that our novel sampling method produces imputing data consistent with the true statistical character as the data, despite using an ``incorrect" noise model in the imputation proposal.}
    \label{fig:pval dist}
\end{figure}

\begin{figure}
    \centering
    \includegraphics[width=1\linewidth]{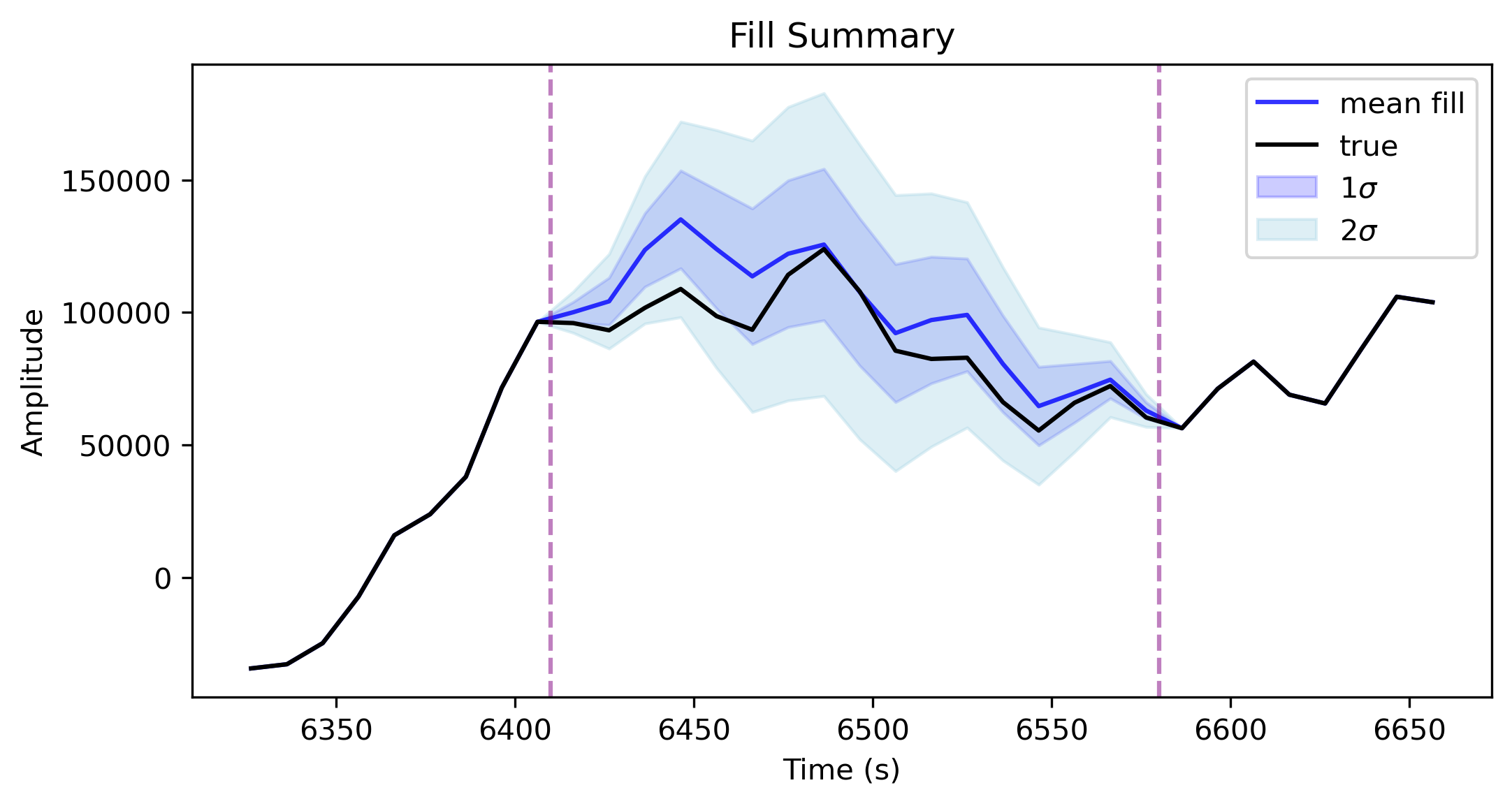}
    \caption{A plot showing the distribution of data imputations in a single gap (gap borders denoted by vertical hashed lines) using this method. The true (missing) data lies within the spread of accepted fills.}
    \label{fig: fill summary}
\end{figure}

We can demonstrate that despite using an ``incorrect" noise covariance in the proposals, this method still produces noise draws that are consistent with the truth. There is also no bias, since in spite of using a prior noise model in the MCMC for the proposal, the likelihood will parse out imputation proposals that are consistent with the current MCMC state. We perform an Anderson-Darling test on data fills generated in this way, whitened with the true noise covariance, to check that it is normally distributed. The p-values recovered from this statistical test for each fill should be uniformly distributed. We performed this test in a toy non-stationary red noise model (see Sec.(\ref{sec: toy model})) for multiple gap scenarios, each to the same general result in Fig.(\ref{fig:pval dist}) and Fig.(\ref{fig: fill summary}). 

It is important to note here that while we are sampling over filling data, no additional parameters are being added to the analysis. The data that is generated for the fills is simply a random draw from the conditional distribution that is immediately subtracted from the observed data in the likelihood.

\subsection{Initialization}
\label{sec: initialization}


One of the challenges any analysis will face is obtaining an initial estimate of the noise PSD. This will have to be done on-source, and in the presence of data corruption. Conceivably, spectral leakage can be mitigated initially with ``primitive" fills, such as simple linear fills, interpolation, or noise draws informed by neighboring uninterrupted data segments. Any of these will likely be effective for short duration gaps, and our preliminary explorations support this. Because the noise in second generation TDI combinations will have a blue spectrum, long scale correlations will be less important, easing the burden of generating initial imputations.

Because drawing a new fill from the imputation proposal distribution is often a large change to the likelihood, acceptance could be low in the burn-in period. To alleviate this, it's possible instead to add other ``small-jump" fill proposals to the repertoire. Conceivable proposals include small random Gaussian adjustments to the points of a previous fill, with variances that taper near gap boundaries to maintain the fill's continuity with the observed data. Or, one could eigen-decompose the conditional distribution's covariance matrix and make fill proposals in one or few of its eigen-directions (this is contrasted with ``normal" fill proposals which make jumps using all eigen-directions).

Once initialized, the more comprehensive augmentation procedure can be implemented. It's also possible to have a dynamically evolving frequency of updates to the imputation proposal distribution, where during the initialization/burn-in period, the proposal is updated at a greater cadence that is later reduced as the MCMC approaches the \textit{maximum a posteriori}. This challenge will be more fully addressed in future work.

\section{Time-Frequency Analysis}
\label{sec: tf analysis}

\subsection{Why Wavelets? \\ (The Problem of Non-Stationary Noise)}


In LIGO, the transient nature of signals is such that the noise does not vary considerably over their duration. Thus, it may be treated as stationary. With no time evolution, the noise covariance matrix, $\bm{\Sigma}_{N\times N}$, in the Fourier domain is diagonal, since the Fourier phases are uncorrelated. This makes the Fourier basis attractive for Bayesian inference in traditional \gw analysis, since the inverse of the covariance matrix appears in the likelihood - in the Fourier domain likelihood calculations are cheap since the covariance matrix costs only $N$ to invert (v.s. $N^3$ for dense matrices). However, the future of \gw detection will push us beyond that useful approximation.


In both ground- and space-based detectors, we expect that the time a signal is observed will increase. In ground-based, this is due to improved low frequency sensitivity, whereby chirping binaries will be seen earlier in their evolution. Space-based detectors naturally target sources which evolve slowly in frequency, and will observe them over the course of months or years (particularly in the case of galactic binaries). In LISA, this timescale is expected to be much larger than that of the noise variation.

Non-stationarity introduces correlations in the Fourier phase which creates off-diagonal elements in the noise covariance matrix, whose size scales with the time series. The now dense and larger covariance matrix makes its inversion far more costly, and more difficult to estimate from the data. With a nominal mission duration of four years, and a modulation from the galactic foreground, instrument state-changes, and glitches, it is certain that LISA's noise will not be stationary. Therefore, it's worth considering an alternate basis which is computationally more tractable. Luckily, a basis which diagonalizes the noise covariance matrix is guaranteed to exist via the Karhunen–Loève theorem~\cite{Stark1994}. More fortunate is the fact a basis exists which also is quite amenable to \gw science - Wilson-Daubechies-Meyer wavelets.

\subsection{Wilson-Daubechies-Meyer Wavelets}


As proposed in earlier work~\cite{Cornish_2020}, we proffer Wilson-Daubechies-Meyer (WDM) wavelets ($\psi_{nm}(t) \in \mathbb{R}$) as a basis. These wavelets form a complete, orthogonal functional basis and are well-localized in both the time and frequency domains~\cite{Necula12}. Working fully in this time-frequency domain provides a few benefits: (i) it diagonalizes the noise covariance matrix for a wide range of non-stationary processes, (ii) it has a fast transform comparable to the FFT, $\mathcal{O}\left(N \mathrm{log}N\right)$, and (iii) binary signals are inexpensive to transform $\mathcal{O}(\sqrt{N})$~\cite{Cornish_2020}.


The WDM wavelet transform of a time series may be represented as a 2D $N_f \times N_t$ grid of pixels, where the value of each pixel is the wavelet amplitude. Each pixel has a uniform extent in time $\Delta T$ and frequency $\Delta F$\footnote{With the exception of the extremal layers in frequency, $m = 0,N_f$. These frequencies span double the temporal extent and half the frequency bandwidth as the ``core" pixels. See Appendix A for details.}, and is labeled with two indices, $\{n,m\} \leftrightarrow \{t_n,f_m\}$.

Each wavelet is constructed by the envelope of a sinusoid. This envelope ``filter" is the Meyer window function which is localized both in time and frequency, with width $K = 2qN_f$ in the time domain (where $q > 1$ is a multiplier setting the size of the window relative to $\Delta T$). The filter length is typically larger than the wavelet pixel temporal extent, $\Delta T$ (with $K_\mathrm{max} = N$). The filter may be further truncated in time arbitrarily for additional speedup with increasing error in the transform. 

The Meyer window is not unique for this time-frequency analysis, any window function may be chosen as long as basis orthogonality is maintained. For example, one could instead choose to ``swap" the time and frequency domain versions of the Meyer window (similar to the tapered window in Mallat \textit{et al.}~\cite{Mallat98}), or use a Gaussian envelope, depending on the desired properties of the basis. We intend to study these window choices and their implications for data augmentation in future work. 


The prime advantage of this wavelet domain is the diagonalization of the noise covariance for a number of non-stationary noise processes, by virtue of their time-frequency localization. Given that noise can be made arbitrarily non-stationary, there is no general proof that the noise covariance will always be diagonal, but it has been shown to be for certain cases relevant to \gw astronomy. Specifically, this may be done for noise spectra that are locally-stationary. Spectra may be considered locally-stationary for time variation timescales larger than the wavelet pixel time span $\Delta T$. In which case, given a dynamic noise spectrum $S = S(f,t)$, the noise covariance in the wavelet domain is,

\begin{equation}
    \Lambda_{(nm)(n'm')} \,\approx\,  S(f_m,t_n)\,\delta_{nn'}\delta_{mm'}\,,
\end{equation}

\noindent which records the covariance between two time-frequency pixels labeled by $\{n,m\}$ and $\{n',m'\}$. The degree to which derivatives in time and frequency of the dynamic noise power spectrum hinder this approximation and necessitate off-diagonal elements is investigated in Ref.~\cite{Cornish2025}. Broadly speaking, if the fractional change in $S(f,t)$ is less than $\sim10\%$ over the time-frequency extent of a wavelet pixel, the off-diagonal elements can be safely neglected. Ref.~\cite{Cornish2025} further discusses strategies for including off-diagonal terms at modest cost when dealing with rapidly-varying noise.


Because the main analysis is performed in the wavelet domain and imputing data is generated in the time domain, we needed to develop the (novel) discrete transformation matrix for WDM wavelets, $\bm{W}_{N\times N}(\mathbb{R})$. The details of its construction, structure, and properties are expounded upon in Appendix A. Suffice to say here, like the discrete Fourier transform, it provides a mapping from the data time series to wavelet amplitudes, $\bm{a}$,

\begin{equation}
    \bm{a} = \bm{W\,d}\,.
\end{equation}

The primary need for the transformation matrix is to move the noise covariance matrix between the time and wavelet domains. The inverse transform (wavelet $\rightarrow$ time domain) of the noise covariance is,

\begin{equation}
    \bm{\Sigma} = \bm{W}^\mathrm{T} \bm{\Lambda} \bm{W}\,.
\end{equation}

\noindent where we have made use of condensed wavelet indices (see Appendix A) to make $\bm{\Lambda}$ a $N \times N$ matrix. 


WDM wavelets are further compelling for their fast transforms, at a cost comparable to an FFT $\mathcal{O}(N\mathrm{log}N)$~\cite{Cornish_2020}, versus the discrete transform at cost $N^2$. For band-limited signals (such as galactic binaries), one may also implement a ``heterodyned" transform, whereby the signal's frequency $f_c$ is base-banded and coarsely sampled to significantly reduce the number of data points in the transform (see section V.A. of Ref.~\cite{Cornish_2020}),

\begin{equation}
    f_c \rightarrow f_c - m_c \Delta F\,,
\end{equation}

\noindent where $m_c = \lfloor{f_c/\Delta F\rfloor}$ is the index of the wavelet frequency layer the signal is in. 


Another advantage of wavelet domain analyses over the Fourier domain is the limited impact of gaps. In the Fourier domain, spectral leakage spreads across all times by virtue of the extent of its basis functions. However in the wavelet domain, the spread of spectral leakage is more compact in time-frequency, since the basis functions are localized. This means that many computations can be cut down. 

In general, transformation via $\bm{W}$ is slow, costing $\mathcal{O}(N^2)$. However in some cases, this transformation may be done quickly for the filling data ($\bm{d}_A = \bm{d}_o + \bm{d}_\mathrm{fill}$),

\begin{equation}
    \bm{d}_\mathrm{fill} = \bm{0}_{N_o}\oplus \bm{d}_i
\end{equation}

\noindent where $\bm{0}_{N_o}$ is the zero-vector of size $N_o$. The transform is,

\begin{equation}
    a_J = \sum_{k \in M} W_{Jk} (d_\mathrm{fill})_k
\end{equation}

\noindent where $M = \{i_m \pm K\}$ is the set of missing (gapped) indices and their surrounding indices affected by the gap, defined by the wavelet filter length $K$. The transformation then becomes $\mathcal{O}(N\times (N_m+2KN_\mathrm{gaps}))$, where $N_\mathrm{gaps}$ is the number of gaps. This speedup is only possible where there are infrequent enough gaps that they are separated by at least a filter length. The transform can be further sped-up. In principle $J$ (a condensed wavelet index, see appendix A) may also be abbreviated in frequency if we wish to perform a band-limited transform, or time, given that the gaps are also infrequent.

Another approach for quickly transforming imputing data in short gaps ($\Delta t_\mathrm{gap} \ll \Delta T$) takes advantage of the time-to-wavelet fast wavelet transform. The standard time to WDM transform of data $\bm{d}$ sampled at cadence $\Delta t$ is given by~\cite{Cornish_2020},

\begin{equation} 
a_{nm} =  \sqrt{2} \Delta t \Re\, C_{nm} X_{n}[mq] \, ,
\end{equation}

\noindent where $C_{nm} = 1$ for $(n+m)$ even, and $C_{nm} = i$ for $(n+m)$ odd, and,

\begin{equation}
X_{n}[j] =  \sum_{k = -K/2}^{K/2-1}  e^{2\pi i k j /K} d[n N_f + k]\phi[k] \,,
\end{equation}

\noindent where $\phi$ is the time domain Meyer window function (see appendices or Ref.~\cite{Cornish_2020}). The $X_n[j]$ can be computed using a FFT at cost $K\ln K$ for each time slice $n$.

\begin{figure}
\includegraphics[width=0.95\linewidth]{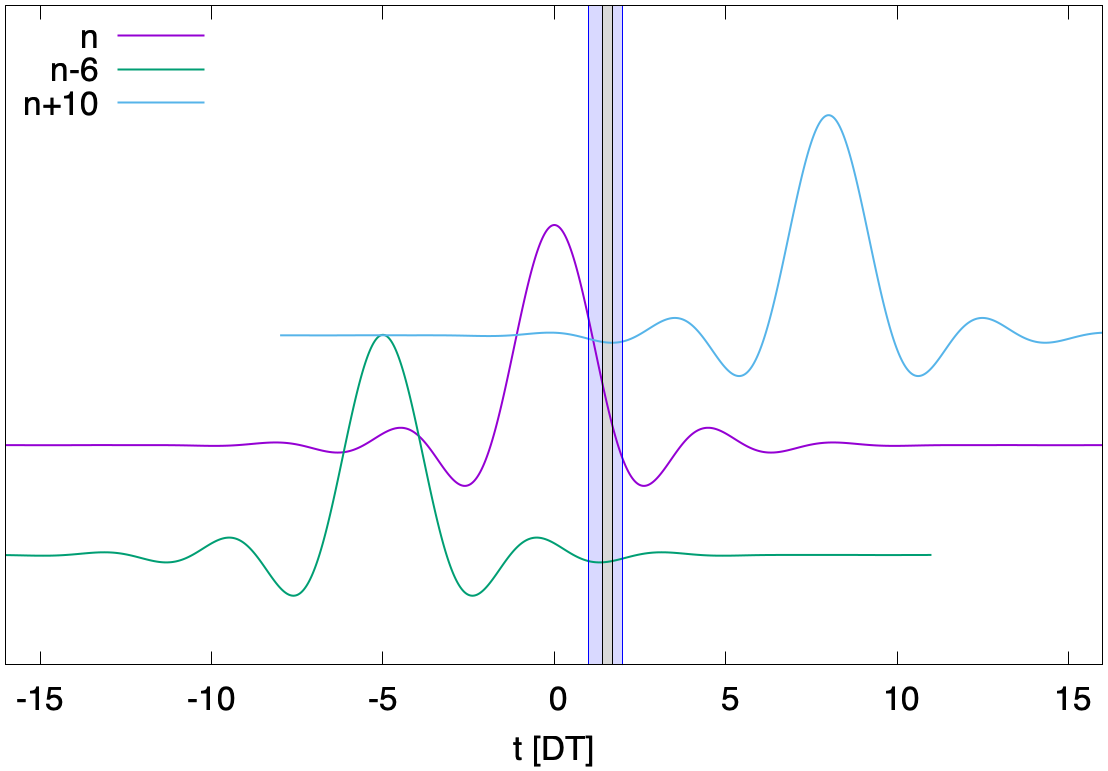}
\caption{Illustration of how the time domain WDM transform can be computed efficiently for short data gaps. In this example, the data gap (grey shaded region) fits inside a single wavelet pixel (blue shaded region). The envelopes of the time-domain wavelet filter functions for three time bands are shown with vertical offsets for clarity. The WDM transform can be computed with no loss of accuracy using an FFT that only uses data in the blue shaded region. The saving is this instance is a factor of $q=16$.}
\label{fig:wdmgap}
\end{figure}

When transforming data that is only non-zero for a duration less than $K\Delta T$, the transform can be sped up by using shorter time spans in the FFT. The savings are equal to $q/p$, where $p$ are the number of wavelet pixels impacted by the gap. The restricted transformation is exact since the data are zero outside of the gap. The coefficients are now read off at $X_{n}[mp]$ rather than  $X_{n}[mq]$. Fig.(\ref{fig:wdmgap}) illustrates the time-restricted WDM transform. In this example, $p=1$, and the computational cost is cut be a factor of $q=16$.

\section{Toy Model}
\label{sec: toy model}

\subsection{Toy noise model}

In our initial explorations and application of the techniques presented in this work we used a simplified noise model (as previously referenced in Sec.(\ref{sec: novel sampling algo})). The noise model has a red frequency spectrum with an amplitude that is modulated in time by a sinusoid, making it non-stationary.
We construct this noise model as the product of two functions which depend on frequency and time, and are shaped by six parameters $\bm{\tau} = \{A,l,\omega_n,\phi_n,s,\alpha\}$ with injected values given in Table~\ref{table: inject params toy}. 

The dynamic spectrum is given by the product
\begin{equation}
    S(f,t; \bm{\tau}) = F(f; \bm{\tau})\,T(t; \bm{\tau})\,,
\end{equation}
\noindent where,
\begin{equation}
    \begin{cases}
        & F(f; \bm{\tau}) = (f^2 + s^2)^{-\alpha/2}\\
        & T(t; \bm{\tau}) = A(1+l\,\mathrm{cos\,(\omega_n t + \phi_n))}\\
    \end{cases}
    \,.
    \label{eq: toy noise}
\end{equation}

\begin{table}[htp]
\caption{Injected values for the toy noise model hyper-parameters.}
\begin{ruledtabular}
\begin{tabular}{c|cdddddd}
Hyper-parameter& Injected Value \\
\hline
\hline
$A$ & $3~\mathrm{Hz}^{-1}$ \\
\hline
$l$ & $0.8$ \\
\hline
$\omega_n$ & $0.15~\mathrm{mHz}$ \\
\hline
$\phi_n$ & $\pi/4$ \\
\hline
$s$ & $1~\mathrm{mHz}$ \\
\hline
$\alpha$ & $4$ \\

\end{tabular}
\end{ruledtabular}
\label{table: inject params toy}
\end{table}

We use this model to illustrate two uses for the wavelet domain Bayesian data imputation: removing edge effects and filling data gaps where the noise model changes across the gap. Note that the noise model parameters listed above, along with an overall change in amplitude across the gap, are explored by the MCMC in conjunction with the gap filling and edge extension.

\subsection{Edge Extension}
Beyond filling gaps, this method of augmentation can also be used to eliminate spectral leakage due to edges on the dataset. In general, we do not expect data to be periodic, so transforming it to the wavelet domain will create spectral leakage at the edges. This may be mitigated by ``extending" the edges with augmented data; there still exists spectral leakage since the imputing data is unconstrained by observed data on the ends and therefore is also not periodic, but the leakage has been ``pushed" out of the region of interest. The benefit of this method is that no data need be lost to remove edge leakage, as would be in apodization. It however is an additional cost, to create more imputing data and transform it to the wavelet domain.

\begin{figure}[htp]
{
  \includegraphics[clip,width=1\columnwidth]{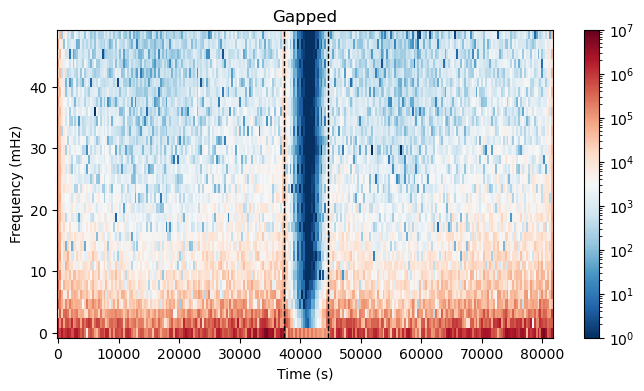}%
}

{
  \includegraphics[clip,width=1\columnwidth]{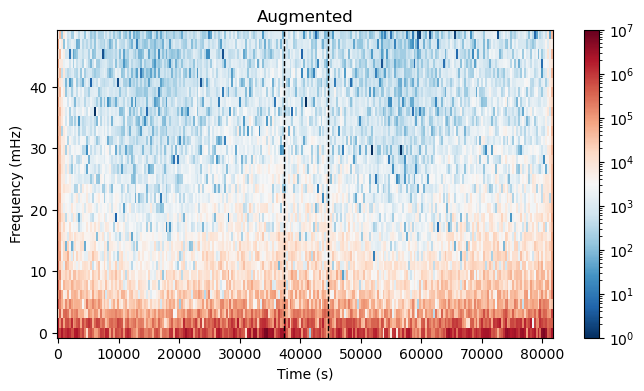}%
}
{
  \includegraphics[clip,width=1\columnwidth]{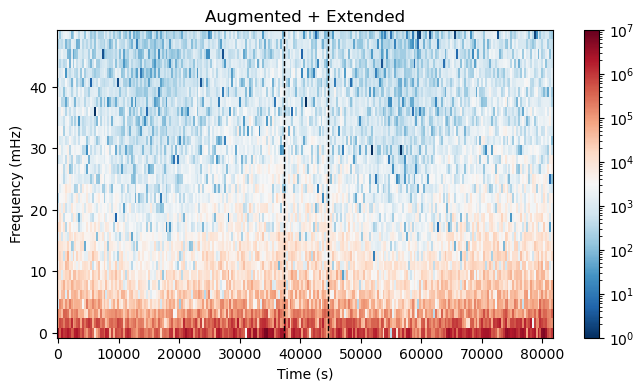}%
}

\caption{Illustration of the edge extension utility in conjunction with a state change in the noise amplitude (increased by 20\%) across the gap in the toy noise model, vertical hashed lines denote the boundaries of a gap. The top panel shows the gapped data with no fill or edge extension.  In the middle panel, a chimeric noise model is used to generate data that smoothly fills the gap and eliminates spectral leakage. In the bottom panel, non-periodic data is padded beyond the observation window with a draw from the conditional distribution, removing spectral leakage at the edges.}

\label{fig: extension}

\end{figure}

To perform this treatment, the data is padded with at most $K$ points on either end, generated using the conditional distribution as in gap filling, Eq.(\ref{eq: conditional dist}) - including noise and signal contributions. An edge impacts surrounding pixels according to the filter length $K$, so that should be the maximal (optimal) choice for the size of padding data. Then after transforming to the time-frequency domain, the padded data can be excluded from the analysis. This is illustrated in Fig.(\ref{fig: extension}).

\subsection{Melding Disjoint Noise Models}
\label{sec: noise melding}

\begin{figure}
    \centering
    \includegraphics[width=1\linewidth]{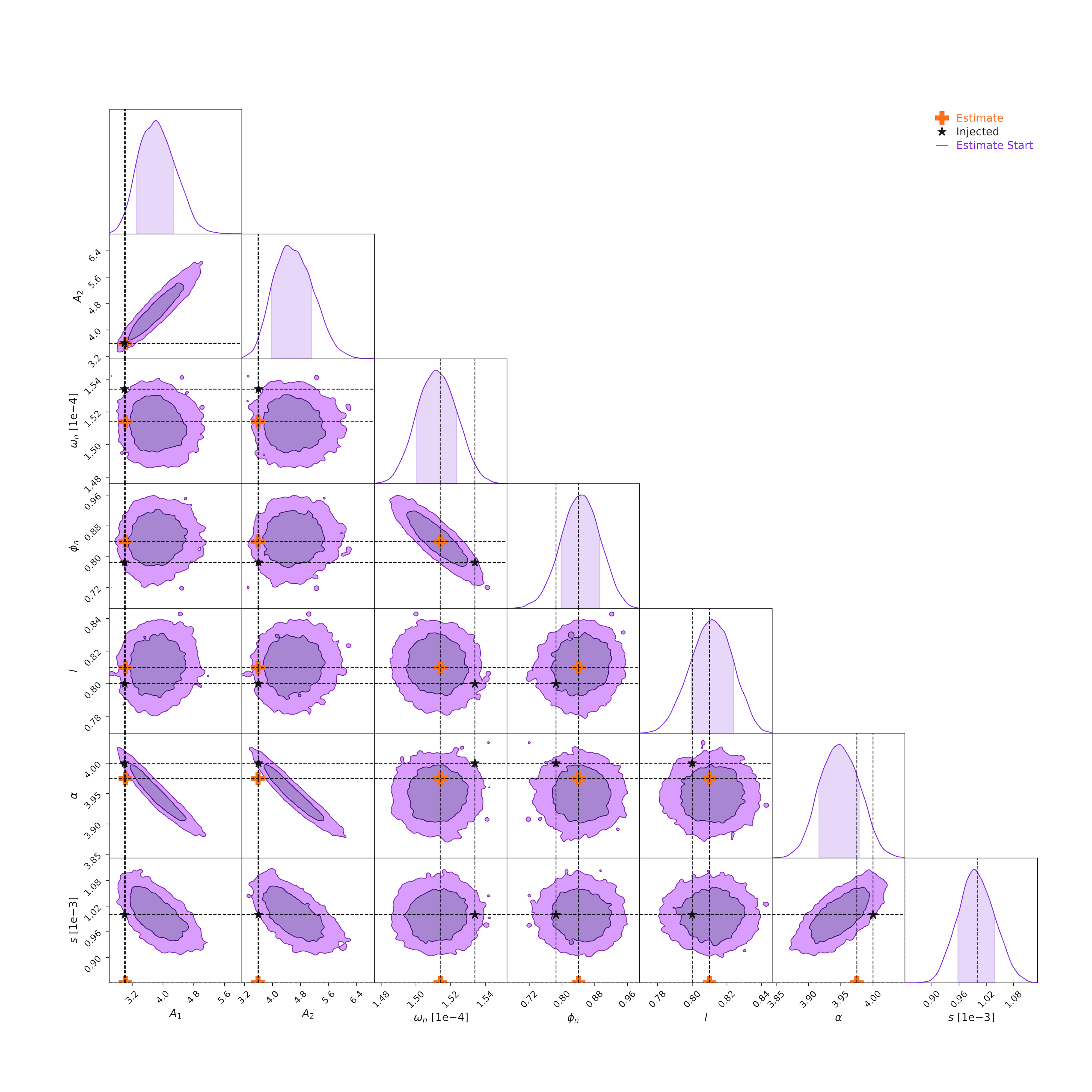}
    \caption{Posteriors of an estimate-initialized run on the toy noise model (post burn-in samples), with a 20\% increase in amplitude ($A_1\rightarrow A_2$) across a single $\sim$ 2 hour gap. The orange cross denotes the location of the estimated parameters, the black star denotes the location of the injected. This demonstrates the algorithm's ability to initialize from an estimate of the noise, and handle state changes in the noise across gaps, and still effectively sample the true posterior.}
    \label{fig: estimate post}
\end{figure}

It is conceivable that the instrument noise will change abruptly after a gap. We show that our method of gap filling can accommodate this state change by melding the noise models, using some smoothly transitioning function between $0$ and $1$. If the noise model changes across a gap $S_1(f,t) = S(f,t; \bm{\tau}_1) \rightarrow S_2(f,t) = S(f,t; \bm{\tau}_2)$, then the chimeric noise model is,

\begin{equation}
    S_\mathrm{chimera}(f,t) = S_1(f,t) w_1(t-t_s) + S_2(f,t)w_2(t-t_s)\,,
\end{equation}

\noindent where $w_1(t),w_2(t)$ are melding functions that smoothly transition to $0,1$ respectively, beginning at the first wavelet pixel the gap is contained in, $t_s$. In our implementation, we use half-Hann windows,

\begin{equation}
    w_1(t) =
    \begin{cases}
         \mathrm{cos}^2\left(\frac{\pi t}{2L}\right)\,, & t \in [t_s,t_e]\\
         1 \,, & t < t_s \\
         0 \,, & t > t_e
    \end{cases}\,,
\end{equation}

\begin{equation}
    w_2(t) =
    \begin{cases}
         \mathrm{cos}^2\left(\frac{\pi (t+L)}{2L}\right)\,, & t \in [t_s,t_e]\\
         0 \,, & t < t_s \\
         1 \,, & t > t_e
    \end{cases}\,,
\end{equation}

\noindent where $L$ is the number of wavelet pixels the gap spans in time, and $t_s,\,t_e$ are the start and end times of pixel time span. We demonstrate the efficacy of this solution by running the toy noise model aforementioned with a 20\% increase in amplitude ($A_1\rightarrow A_2$) across a single $\sim$ 2 hour gap, in conjunction with the edge extension utility, see Fig.(\ref{fig: estimate post}). In this example we set $N = 8,192$, $N_\mathrm{gap} = 720$, and $f_s = 0.1$ Hz. The sampler was initialized with an estimate discussed in the following section. In this case with a disjoint noise model, the edges were extended with the corresponding noise model before and after the gap, i.e. the edge prior to the gap was extended using $S_1(f,t)$ and the edge after the gap with $S_2(f,t).$ Were there a signal, its parameters and contribution to the conditional distribution would be identical either side of a gap, since the instrument's behavior is independent of the signal source.

This approach of making a chimeric noise model can become difficult as the gap becomes shorter in duration and the difference in noise models across the gap becomes greater. This strains the assumption in using our wavelet basis that the noise is locally stationary - that it doesn't vary significantly over a wavelet pixel time span, $\Delta T$. A solution to this possibility is to sever the analysis. The data prior to the gap is analyzed with one noise model and the data following the gap with another model. The gap then effectively becomes two edges that can be extended using their respective models. 

In reality, we will not know \textit{a priori} whether or not there is a noise state change across any gap. The most likely approach to this problem is to model the noise agnostically by allowing the noise model parameters to be different in each segment of data as demonstrated in our toy model. The initialization process could proceed as described in Sec.(\ref{sec: initialization}). A full implementation of this procedure will be explored in future work.

\subsection{Initialization Demonstration}
\label{sec: init demo}

To explore the chimeric noise model we need to initialize the sampler at some point in parameter space. We briefly discussed initialization in Sec.(\ref{sec: initialization}), but here we will demonstrate the robustness of the algorithm to the starting point of the noise model. There are many well-established techniques for estimating the location of posterior modes. In our prototype implementation, we are using a simple single-chain MCMC algorithm which is sufficient for sampling around the peak of the posterior, but is not well suited to finding the peak. In future implementations, we intend to make fuller use the tools of such as parallel tempering and maximum likelihood estimates to more effectively burn-in to the injection~\cite{Cornish2014,Littenberg2023,Gupta2023}. 

To obtain a decent starting point for the sampler, we use classical Welch-like estimates and a maximum likelihood solution. To estimate the spectral component of the toy noise model $\{\alpha,s\}$, we applied a rolling half-sine window to the time series, advancing the window each step by half its width. Its width was set by the desired frequency resolution $\sim$mHz. In each windowed segment, we obtained a periodigram and maximized the Whittle log-likelihood to estimate the overall amplitude in each segment, $T(t)$, the spectral slope $\alpha$, and the knee frequency $s$. The Whittle log-likelihood for each segement is given by~\cite{Gupta2023},
\begin{equation}
    \mathrm{ln}\mathcal{L} = -\sum_{i = 1}^{M/2}\,\mathrm{ln}\left[\frac{T(t)}{(f_i^2+s^2)^{\alpha/2}}\right] + \frac{P(f_i)(f_i^2+s^2)^{\alpha/2}}{T(t)}\,,
\end{equation}
\noindent where $P(f)$ is the periodigram found in a windowed segment, and $M$ is the number of data points the segment. The log-likelihood can be analytically maximized with respect to $T(t)$, while maximizing with respect to $\{\alpha,s\}$ was done using a grid search for $\alpha\in[3.5,4.5],\,s\in[0.5,1.5]\,\mathrm{mHz}$ with 20 points in each interval.
Since we know that the spectral shape is fixed in this example, an aggregate estimate for $\{\alpha,s\}$ was found by averaging the maximum likelihood solutions from each segment.
Note that we did exclude segments that overlapped with the gap.
The maximum likelihood solution for $T(t)$ in each segment was recomputed using the aggregate values for $\{\alpha,s\}$ given by $s=0.8403,\,\alpha=3.9750$.

\begin{figure}
    \centering
    \includegraphics[width=1\linewidth]{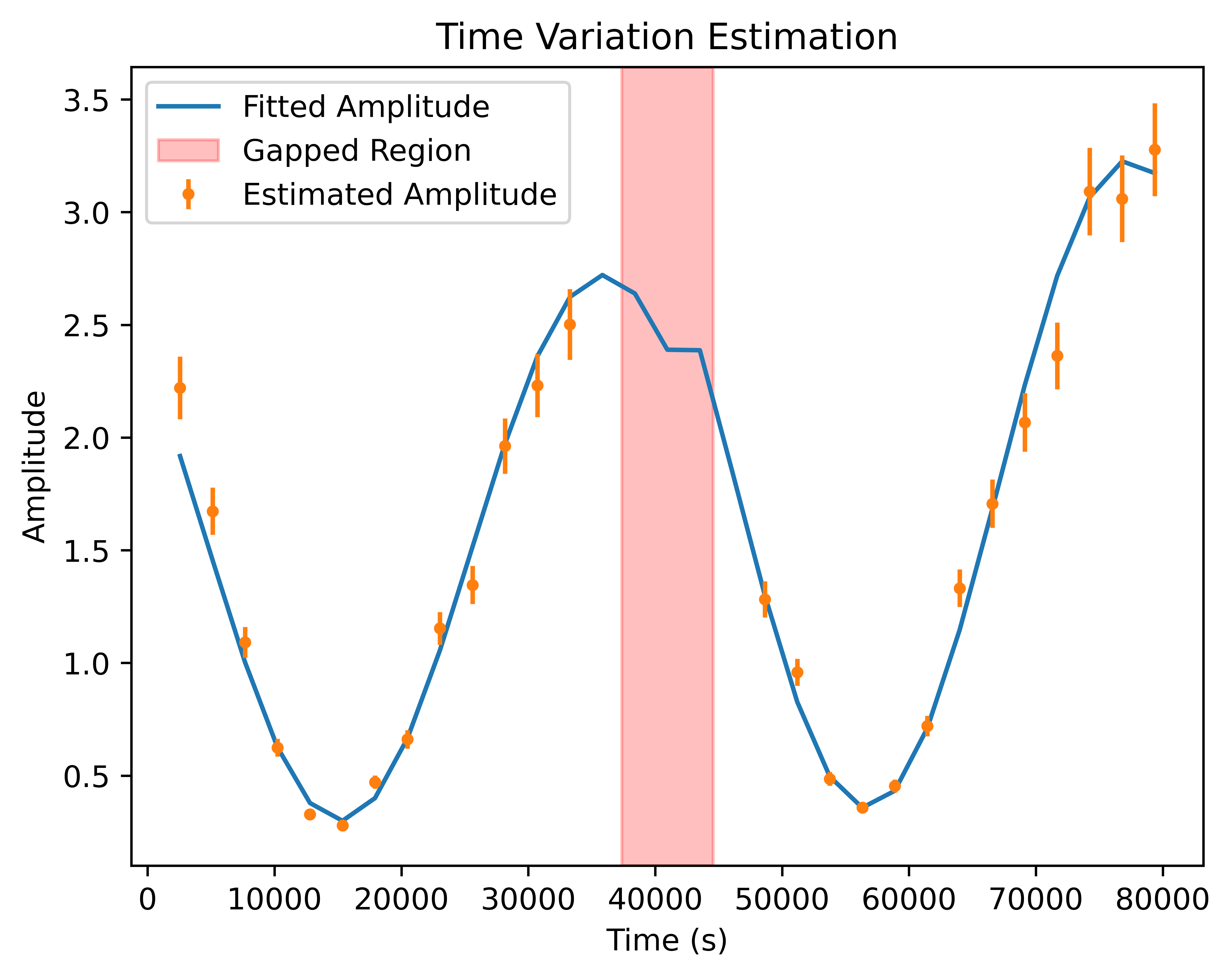}
    \caption{Maximum likelihood estimate of the time modulation of the toy noise model with an amplitude jump across the gap. Vertical bars on the estimated amplitudes are Fisher errors.}
    \label{fig: tvar estimate}
\end{figure}

For the time variation component of the model $\{A_1,A_2,l,\omega_n,\phi_n\}$, we minimized the chi-squared,
\begin{equation}
    \chi^2 = \sum_{\mathrm{i = \,segments}} \left(\frac{A_i^\mathrm{estimate}-B_i^\mathrm{model}}{\sigma_i}\right)^2\,,
\end{equation}
\noindent where $A_i^\mathrm{estimate}$ is the estimated amplitude in each windowed segment, $B_i^\mathrm{model}$ is the model $T(t)$ in each segment \textit{convolved} with the square of the window function, and $\sigma_i$ is the Fisher error of the amplitude estimate. The fractional error in $T(t)$ is given by the Fisher information matrix, and scales as $(M/2)^{-1/2}$. Note that the sum excludes segments whose window overlaps with the gap. A grid search in the parameters $\{A_1,A_2,l,\omega_n,\phi_n\}$ was used to minimize the chi-squared. This grid search yields the estimate for the injection: $\{A_1 = 3.0200,\,A_2 = 3.5877,\,l=0.8100,\,\omega_n = 0.1514,\,\phi_n = 0.8400\}$ in their respective units. The results of this approach are plotted in Fig.(\ref{fig: tvar estimate}). 

We then initialized the sampler with this estimate and produced the posterior distribution shown in Fig.(\ref{fig: estimate post}). Note that the initial guess lies close to the injected parameters, falling within the posterior distributions for most of the parameters.

In future work, we intend to further develop the initialization techniques. One could imagine that the estimate could be refined by filling the gaps via data augmentation using the initial estimate, which in turn could be used to produce a new estimate (although in the sampling algorithm this essentially happens once it's initialized). We also don't address here the presence of a signal and leave its full treatment to future work with more sophisticated tools.

\subsection{Injection-Recovery Verification}
\label{sec: pp plotting}

The posteriors presented in Fig.(\ref{fig: estimate post}) do not show any obvious biases, but this is for one noise realization. To demonstrate more concretely that this analysis is bias-free, we generated a Probability-Probability (PP) plot, a standard demonstration in Bayesian parameter inference~\cite{Cook2006} (see Refs.~\cite{Gair2022,Wouters2024} for additional notes on generating PP plots). To do so, we re-ran the toy model analysis 100 times, with each simulated noise realization generated with a new parameter draw from the same priors as the MCMC sampling ($A_2$ was also a fair draw, with an identical prior as $A_1$). We placed an additional constraint that the draws must produce a positive definite conditional covariance matrix (this is the same constraint that naturally appears in the sampling). After performing this batch of analyses, we calculated the credible level at which each injected parameter appears in its marginal posterior. This is done by ordering the samples in each run for each parameter in ascending order, then calculating the empirical cumulative distribution function, and finding the credible level of the injected parameter. 

\begin{figure}
    \centering
    \includegraphics[width=1\linewidth]{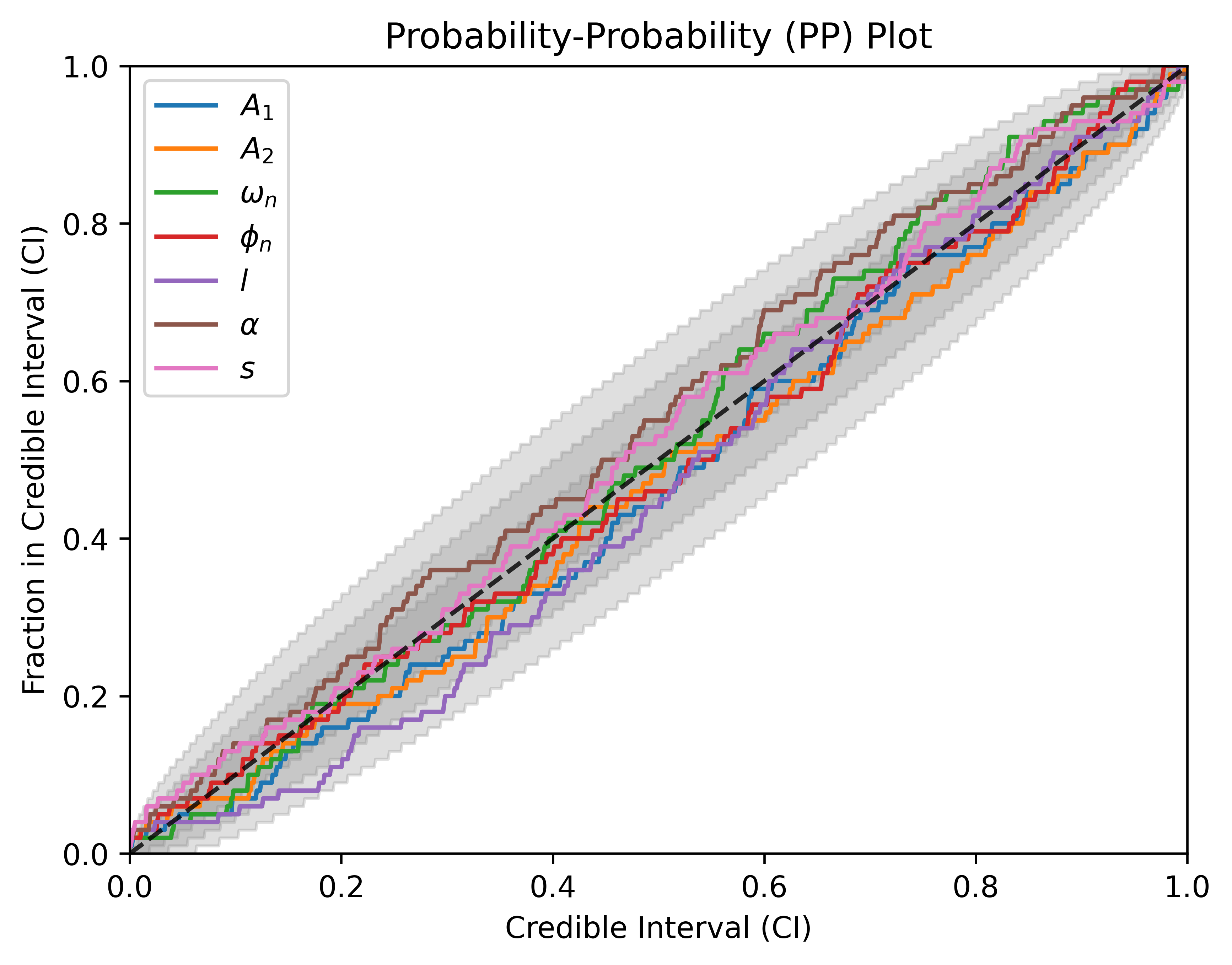}
    \caption{Probability-Probability (PP) plot generated from multiple runs of the toy model analysis, qualitatively showing no evidence of a bias in injected parameter recovery. The gray shaded regions show the $1\sigma,2\sigma,$ and $3\sigma$ confidence levels due to a finite number of runs, the diagonal dashed line shows the ideal.}
    \label{fig: pp plot}
\end{figure}

If the analysis is bias-free, we should expect that each injected parameter appears in the $x\%$ credible interval of its marginal posterior $x\%$ of the total number of runs. Thus, in the ideal scenario plotting this fraction versus the credible interval should yield a plot that follows the diagonal. Because we perform a finite number of runs, there is expected to be variation about this ideal, which is given by the inverse cumulative distribution function (the quantile function) of the binomial distribution (where $n$ is the number of runs, $p$ is the credible level, and $q$ is the confidence quantile). In the PP plot we produced for the toy model, Fig.(\ref{fig: pp plot}), we qualitatively see results consistent with a bias-free analysis. 

\section{LISA-Like Example}
\label{sec:simulation}


\subsection{Noise and Signal Models}
\label{sec:noise&sig}

To demonstrate the effectiveness of this tool, we simulated one year of LISA data, as observed from one Michelson-style interferometer channel. The noise has stationary and non-stationary components from the instrument and galaxy respectively, and a simulated non-chirping galactic binary embedded within it.

\subsubsection{LISA-Like Noise Model}

For an effective noise model that approximates the LISA observatory, we used the LISA sensitivity curve, overlaid by a time-varying galactic background~\cite{Robson2018,Digman2022},

\begin{equation}
    S(f,t; \bm{\tau}) = S_n(f; \bm{\tau}_n) + S_c(f; \bm{\tau}_c)\,r(t; \bm{\tau}_m)\,,
\end{equation}

\noindent characterized by eleven parameters $\bm{\tau} = \{\bm{\tau} _n,\bm{\tau}_c,\bm{\tau}_m\} = \{A_o,A_a,A,\alpha,f_1,f_2,f_\mathrm{knee},A_1,A_2,\phi_1,\phi_2\}$. These parameters characterize the contribution to the noise from the instrument, galactic foreground, and its time modulation. See Table~\ref{table: inject params} for the values used to generate the simulated LISA data.

\begin{figure}
    \centering
    \includegraphics[width=1\linewidth]{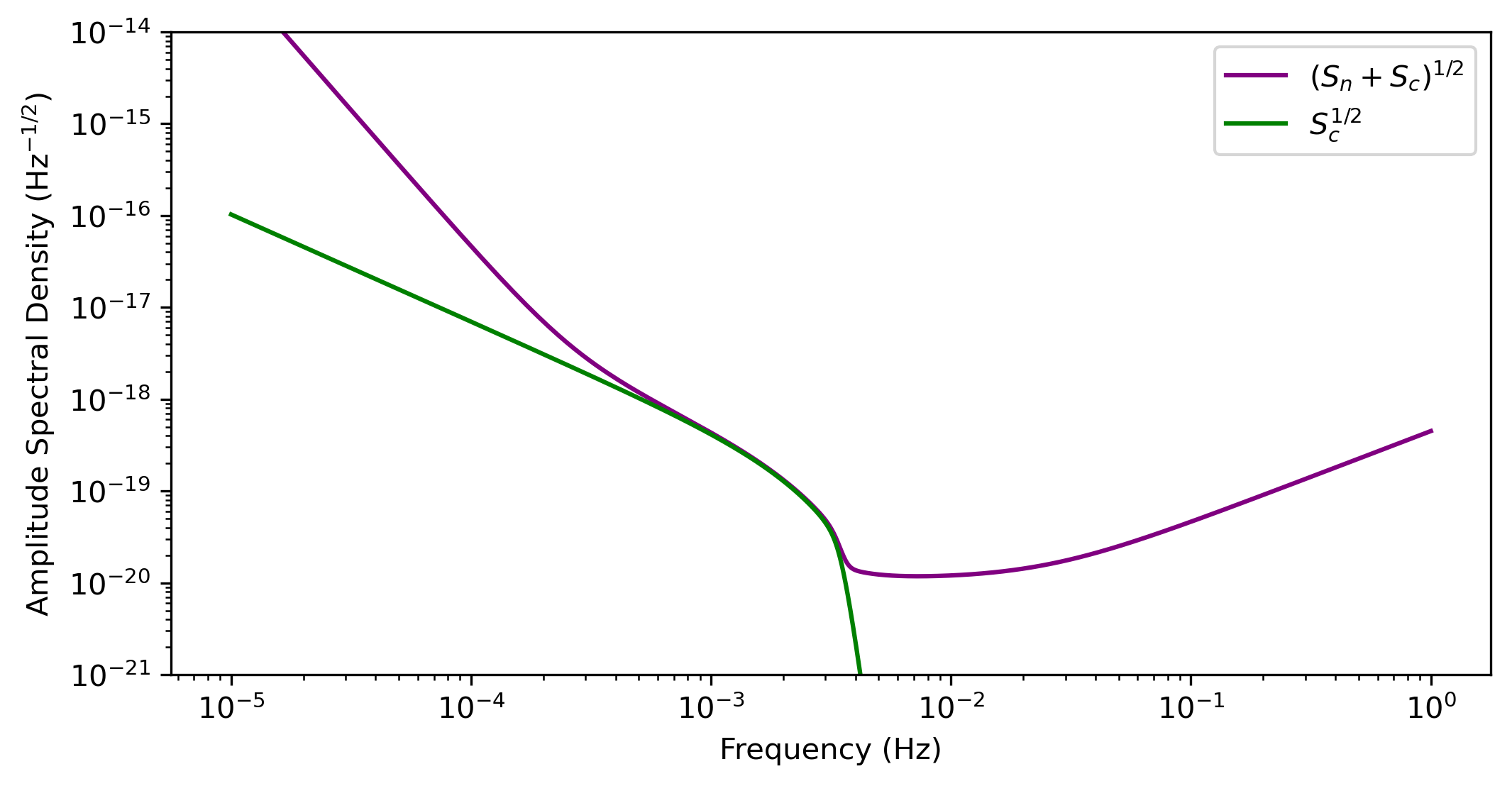}
    \caption{Our time-averaged noise model ($\overline{r}(t)=1$), with contributions from the instrument $S_n(f)$ and galaxy $S_c(f)$.}
    \label{fig: noise model f}
\end{figure}

\begin{table}[b]
\caption{Injected values for the noise model hyper-parameters. These are for one year of observed data, sampled at a frequency $f_s \approx 8\,\textrm{mHz}$.}
\begin{ruledtabular}
\begin{tabular}{c|cddddddddddd}
Hyper-parameter& Injected Value \\
\hline
\hline
$A_o$ & $1.5\times10^{-11}\,\mathrm{m}$ \\
\hline
$A_a$ & $3\times10^{-15}\,\mathrm{m\,s}^{-2}$ \\
\hline
$A_c$ & $1.14\times10^{-44}$ \\
\hline
$\alpha$ & $1.80$ \\
\hline
$f_1$ & $1.99\,\mathrm{mHz}$ \\
\hline
$f_2$ & $0.31\,\mathrm{mHz}$ \\
\hline
$f_\mathrm{knee}$ & $3.38\,\mathrm{mHz}$ \\
\hline
$A_1$ & $0.183$ \\
\hline
$A_2$ & $0.616$ \\
\hline
$\phi_1$ & $3.92$ \\
\hline
$\phi_2$ & $3.09$ \\
\end{tabular}
\end{ruledtabular}
\label{table: inject params}
\end{table}

\begin{figure}
    \centering
    \includegraphics[width=1\linewidth]{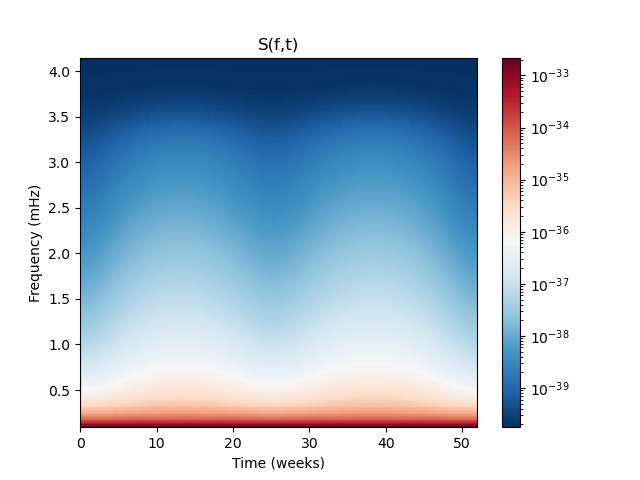}
    \caption{Our noise model $S(f,t)$ in the time-frequency domain. The contribution from the galaxy makes it non-stationary.}
    \label{fig: noise model tf}
\end{figure}

For simplicity, we assume the instrument noise $S_n(f)$ is stationary. The galactic foreground $S_c(f)$, comprised of unresolved galactic binaries, is however non-stationary since it is modulated in time as the detector's antenna pattern rotates relative to the galactic center (approximated by the product with a scaling function $r(t)$~\cite{Digman2022}. We chose the scaling function for the `A' Michelson channel). The spectrum is depicted in the frequency domain in Fig.(\ref{fig: noise model f}) and in the time-frequency domain in Fig.(\ref{fig: noise model tf}). Over the millihertz band, the noise is clearly time-varying and spectrally red.

\noindent For an effective model of LISA's instrument noise $S_n(f)$, we chose to use the LISA sensitivity curve, parameterized with two amplitude parameters, $\bm{\tau}_n = \{A_o,A_a\}$~\cite{Robson2018},

\begin{eqnarray}
    S_n(f;\bm{\tau}_n) &=& \frac{10}{3L^2}\left(P_\mathrm{OMS}(f;A_o)+\frac{4P_\mathrm{acc}(f;A_a)}{(2\pi f)^4}\right) \nonumber \\
    && \times \left(1+\frac{6}{10}\left(\frac{f}{f_*}\right)^2\right)\,,
\end{eqnarray}

\noindent with $L = 2.5\,\mathrm{Gm}$ and $f_* = 19.09\,\mathrm{mHz}$. $P_\mathrm{OMS}(f)$ is the single-link optical metrology noise,

\begin{equation}
    P_\mathrm{OMS}(f;A_o) = A_o^2\left(1+\left(\frac{2\,\mathrm{mHz}}{f}\right)^4\right)\,\mathrm{Hz}^{-1}\,,
\end{equation}

\noindent where the injected value is $A_o = 1.5\times10^{-11}\,\mathrm{m}$.

\noindent and $P_\mathrm{acc}(f)$ is the single test mass acceleration noise,

\begin{eqnarray}
    P_\mathrm{acc}(f;A_a) &=&A_a^2\left(1+\left(\frac{0.4\,\mathrm{mHz}}{f}\right)^2\right) \nonumber \\
    && \times \left(1+\left(\frac{f}{8\,\mathrm{mHz}}\right)^4\right)\,\mathrm{Hz}^{-1}\,.
\end{eqnarray}

\noindent where the injected value is $A_a = 3\times10^{-15}\,\mathrm{m\,s}^{-2}$.

The instrument noise is overlaid by the contribution from the galaxy, characterized by five parameters, $\bm{\tau} = \{A_c,\alpha,f_1,f_2,f_\mathrm{knee}\}$~\cite{Karnesis2021},

\begin{equation}
    S_c(f;\bm{\tau}_c) = \frac{A_c}{2} f^{-7/3}\mathrm{e}^{(f/f_1)^\alpha}\left[1+\mathrm{tanh}((f_\mathrm{knee}-f)/f_2)\right]\,.
\end{equation}

\begin{figure}
    \centering
    \includegraphics[width=1\linewidth]{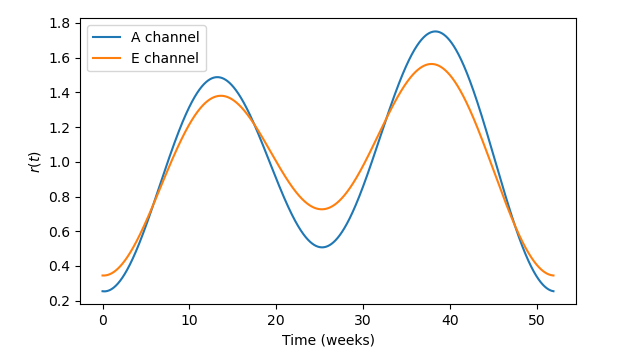}
    \caption{The time modulation function $r(t)$ over one year, for the A and E channels, with fit parameters for the A channel given in Table~\ref{table: inject params}.}
    \label{fig:time mod}
\end{figure}

\noindent The time variation of the galactic background may be approximated by the product with a modulating function with four parameters, $\bm{\tau}_m = \{A_1,A_2,\phi_1,\phi_2\}$~\cite{Digman2022},

\begin{equation}
    r(t; \bm{\tau}_m) \approx 1+\sum_{k=1}^2 A_k\,\mathrm{cos}\,(2\pi tk/T -\phi_k)\,,
\end{equation}

\noindent where we choose to only use the two dominant modes $k = 1,2$, and choose to model Channel A, see Fig.(\ref{fig:time mod}). In practice, the time modulation doesn't need to be fitted, rather shape parameters of the Milky Way provide a mapping to these parameters.

Data may be generated in the time domain by Cholesky-decomposing the time-domain covariance matrix $\bm{\Sigma}(\bm{\tau}) = \bm{HH}^\mathrm{T}$ defined by our noise model $S(f,t; \bm{\tau})$,

\begin{equation}
    \bm{n} = \bm{H\,\gamma}\,,
\end{equation}

\noindent where $\bm{\gamma}\sim\mathcal{N}(0,1)$ is a $N\times1$ vector of Gaussian white noise. Alternatively, noise may be generated in the time-frequency (wavelet) domain directly,

\begin{equation}
    n(f_m,t_n) = \sqrt{S(f_m,t_n)}\,\Upsilon_{nm}
\end{equation}

\noindent where $\bm{\Upsilon}\sim\mathcal{N}(0,1)$ is a $N_f\times N_t$ matrix of Gaussian white noise.

\subsubsection{Signal Model}

As a rudimentary signal model, which effectively mimics the behavior of galactic binaries (with negligible frequency derivatives), we use a simple monochromatic sinusoid described by three parameters, $\bm{\theta} = \{A_s,\omega_s,\phi_s\}$,

\begin{equation}
    h(t;\bm{\theta}) = A_s\,\mathrm{sin}\,(\omega_s t + \phi_s)
\end{equation}

For simplicity in this initial application, we only inject one binary ($\mathrm{SNR} \sim 35$), embedded within the frequency band of the detector subject to galactic foreground modulation.The injected values for the signal parameters are given in Table (\ref{table: inject signal params}). Full global analyses on more realistic LISA data with multiple binaries require trans-dimensional MCMC's~\cite{Littenberg2023}.

\begin{table}[b]
\caption{Injected values for the signal model parameters. These are for one year of observed data, sampled at a frequency $f_s \approx 8\,\textrm{mHz}$, with an SNR $\sim$ 35.}
\begin{ruledtabular}
\begin{tabular}{c|cddd}
Signal Parameter & Injected Value \\
\hline
\hline
$A_s$ & $1\times10^{-20}$ \\
\hline
$\omega_s$ & $f_s/4\approx 2\,\mathrm{mHz}$ \\
\hline
$\phi_s$ & $\pi/4$ \\
\end{tabular}
\end{ruledtabular}
\label{table: inject signal params}
\end{table}

\subsection{Gap Simulation}

\begin{figure}[h]
    \centering
    \includegraphics[width=1.1\linewidth]{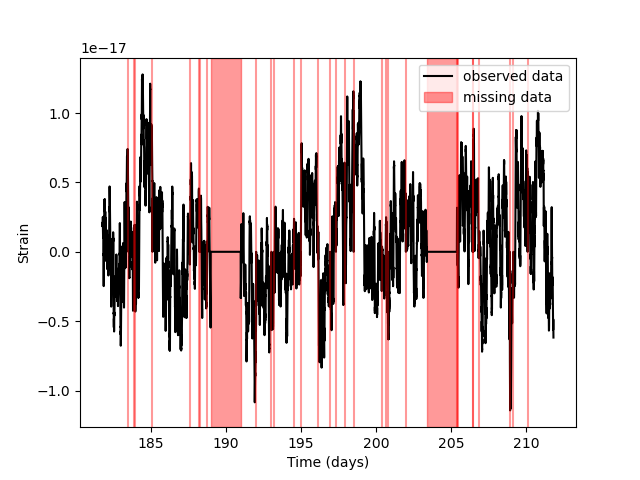}
    \caption{A one-month duration segment of the simulated time series, showing the density of daily short duration random gaps, and scheduled long duration gaps.}
    \label{fig:gapped time series}
\end{figure}

In principle, this method works regardless of the frequency of occurrence or duration of gaps. For the application of this prototype, we produce a pattern of gaps faithful to the forecasts of noise corruption expected in LISA data, as described in Sec.(\ref{sec: data gaps}). There is a mixture of short- and long-duration gaps with a similar frequency of occurrence (see Fig.(\ref{fig:gapped time series})). 

Short gaps are taken to last on average ca. 4 minutes each, and long gaps 2 days each (spanning $<1\Delta T$ and $6\Delta T$, respectively). Short gaps occur with a frequency of roughly once per day, and long gaps occur on a periodic schedule of once per two weeks. Short gaps were uniformly randomly placed.

\subsection{Sampler Details}

To demonstrate this tool, we performed joint parameter estimation on the noise and signal models through blocked sampling with a single-chain Metropolis-Hastings MCMC, with simple proposals from the Fisher information matrix, and Differential Evolution. The sampler state is initialized with the injected parameters, simulating the post burn-in exploration. The frequency of updates to the Fisher matrix, the moments of the conditional distribution Eq.(\ref{eq: conditional dist}), noise imputation, and signal imputation can be independently varied. The code which performs this analysis will be made publicly available\footnote{https://github.com/NoahPearson/Data-Augmentation}.

\subsection{Computational Speedups and Cost}

While the augmentation method we have outlined above is more computationally manageable than previous applications, and we have fast transforms between the time and wavelet domains (which are only necessary for the imputations, which are generated exclusively in the time domain), we'd like to introduce all possible accelerations into the analysis. 

A simple way to introduce a speed-up to this algorithm is to manage the frequency with which various updates to the conditional distribution Eq.(\ref{eq: conditional dist}) and imputing data are made. One must however be wary of changing the frequency of updates to the signal filling - since having a signal fill that doesn't match the current state of the signal sampler will introduce a bias. However, for updating frequencies close to one, the difference between the state of the signal sampler and the signal fill is tantamount to a noise perturbation, and thus does not pose a significant source of bias. 

In the implementation of updating the conditional distribution, rather than multiply masking matrices with the noise covariance (e.g. as in $\bm{\Sigma}_{oo}$) one may instead delete appropriate rows and columns from the full noise covariance. $\bm{M}_o$ effectively deletes missing indices and $\bm{M}_m$ deletes observed indices.

As noted in Baghi \textit{et al.}~\cite{Baghi_2019}, updates to the conditional distribution's moments Eqs.(\ref{eq: conditional dist}) are intensive since they require the inversion of a dense, large matrix $\bm{\Sigma}_{oo}$ nearly the size of the data ($N_o \sim N)$. For realistic simulations of LISA where $N\sim 10^{5-6}$, this is impractical on most PC's. The authors chose to assume that correlations are short-scale (i.e. a fast-decaying autocovariance function), and limit the covariance calculation to a neighborhood of points around each gap, thus performing the augmentation for each gap individually. This also neglects inter-gap correlations, but allows for more sophisticated augmentation techniques like updating gap fills with varying cadences. Because we work in a localized basis, we can make this approximation more naturally, since the impact of a gap extends only as far as the wavelet filter length $K$, this makes for a natural choice of the size of the neighborhood. Further, while in the previous work by Baghi \textit{et al.} this neighborhood approximation to the covariance is effectively an approximation on the \textit{posterior}, here it is only an approximation on the \textit{proposal}. We then can reduce the computer memory required, and perform imputations more quickly. Recall that since we only use this distribution as a proposal, making approximations or simplifications to computing it should only affect its efficiency, not the accuracy of imputations. One then may consider further reducing the size of the neighborhood, just to the point where efficiency becomes a hindrance.

In practice, the matrix products required to compute the neighborhood time domain covariance $\tilde{\bm{\Sigma}}$ are typically of order $N_f^2$.

\begin{equation}
    \tilde{\Sigma}_{ij} = (W^\mathrm{T})_{iJ}\Lambda_{JL}W_{Lj}\,,
\end{equation}

\noindent $i,j$ run over the time domain indices of the neighborhood around the gap, and $J,L$ run over the condensed wavelet indices that contain the neighborhood (the number of indices will be some multiple of $N_f$, less than or equal to $2q$). Since $\bm{\Lambda}$ is diagonal, its product is the trivial multiplication of the rows of $\bm{W}$ by its respective diagonal element. Then the matrix product reduces to the product of two $\mathcal{O}(N_f \times N_f)$ or smaller matrices.

In practice, this becomes complicated for frequent, closely-spaced gaps that occur within a $K$-neighborhood of each other. Closely-spaced gaps effectively reduce the number of neighboring points that condition the imputing data. One could choose to increase the size of the neighborhood to include multiple gaps and fill them simultaneously, or maintain the individual fill procedure with less-conditioned fills. Once the gaps have been filled once in an MCMC, closely-spaced gaps are no longer as problematic, since one could treat the imputing data of neighboring gaps as ``observed" points; the philosophy is that even a poor fill is better than nothing to condition the imputing data. 

\begin{figure}
    \centering
    \includegraphics[width=1.1\linewidth]{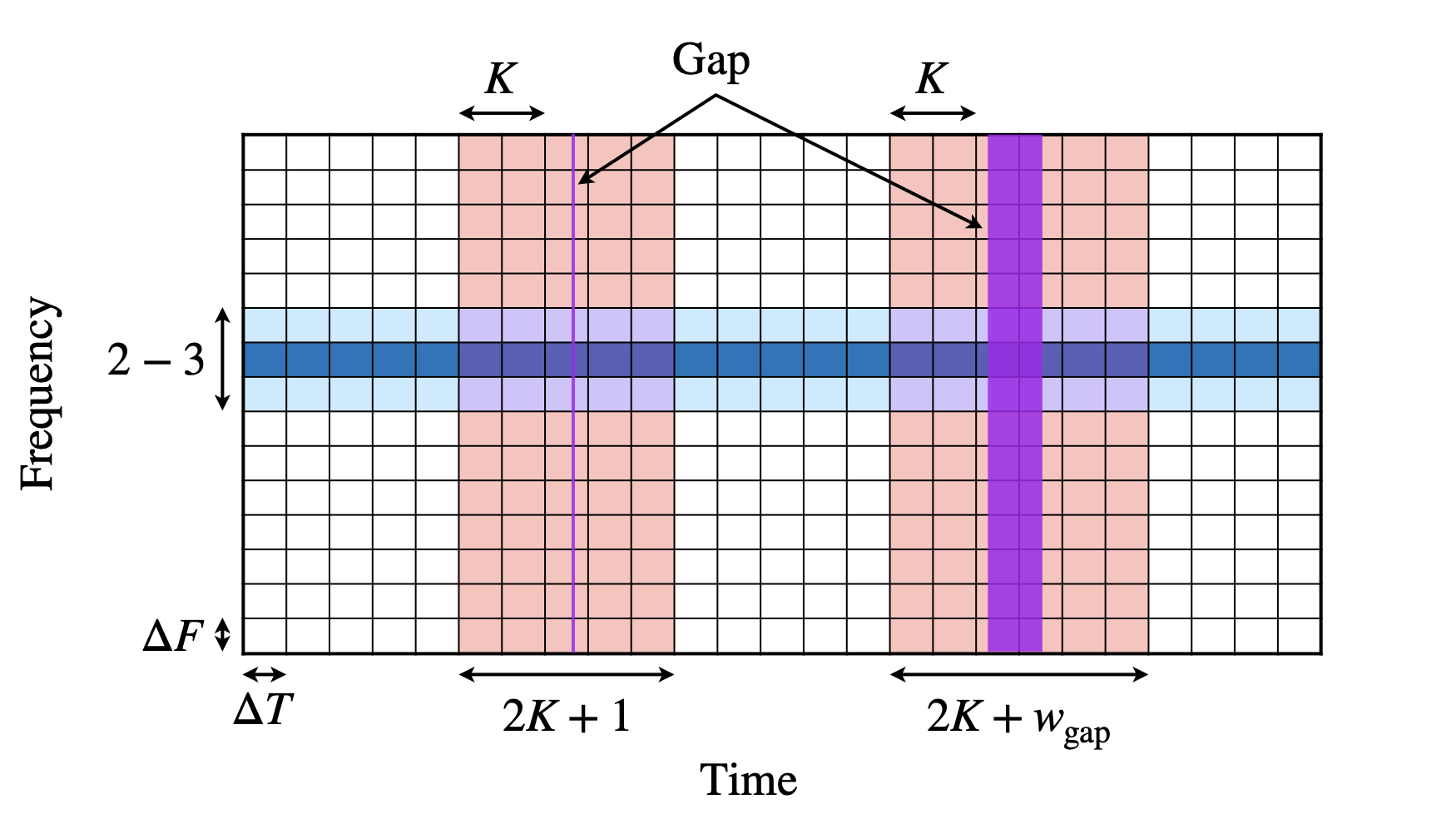}
    \caption{Cartoon illustrating the hierarchy of time-frequency regions of relevance for calculating the likelihood. The most expensive likelihood update is the noise model which spans the whole domain, followed by updates to the noise imputation, and lastly updates to the signal and signal imputation updates.}
    \label{fig:tf likelihood}
\end{figure}

Again, thanks to the time-frequency localization of our basis, we can cut down on costs on updates to the likelihood. Since the impact of gaps and their imputing data extend as far as the transform kernel, the likelihood need only be updated in that span in time. Likewise, band-limited galactic binaries only occupy at most 3-4 frequency pixels. This means that their likelihoods and filling updates to the likelihood can be severely limited in time-frequency. The hierarchy of costs is visualized in Fig.(\ref{fig:tf likelihood}). By far the most expensive likelihood comes from updates to the noise model, which spans the whole time-frequency domain, followed by updates to the noise imputation, and lastly updates to the signal and signal imputation updates.

\section{Results}
\label{sec: results}

\subsection{Application to Simulated LISA Data}

To present an illustration of this tool, we apply it to simulated LISA-like Michelson-style data with realistic gaps. The models for the noise and signal are described in the previous section. 

We choose to analyze one year of data, sampled at $\approx$ 8 mHz, with a time-frequency resolution of $N_t = 1,024,\,N_f = 256$, equating to a data size of $N = 262,144$. With a factor $q = 8$, the wavelet kernel $K \approx 5$ days. The sampler was run for $1\times10^6$ iterations, comprising 10 alternating sampling blocks of the noise and signal models. Updates to the noise imputation were 1/10,000 iterations, updates to the imputation proposal distribution 1/10,000 iterations, and updates to the signal imputation were 1/50 iterations. Because this is a blocked sampler (alternating sampling periods between the signal and noise models) we may infrequently update the filling noise without incurring bias. While the sampling rate is much lower than that of current data challenges, one could achieve the same rate by resampling the data set or band-passing; the choice here was to simply limit the computational cost in this prototype implementation.

\begin{figure}[htp]
{
  \includegraphics[clip,width=0.7\columnwidth]{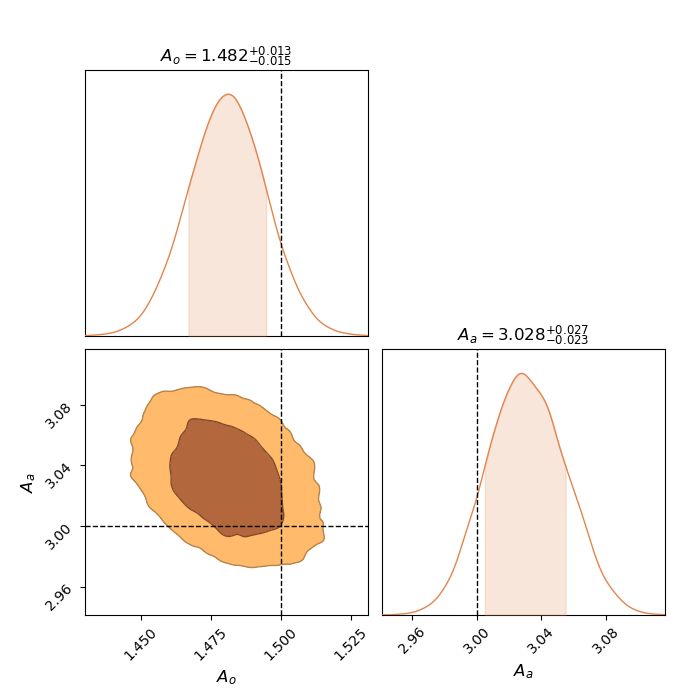}%
}

{
  \includegraphics[clip,width=0.9\columnwidth]{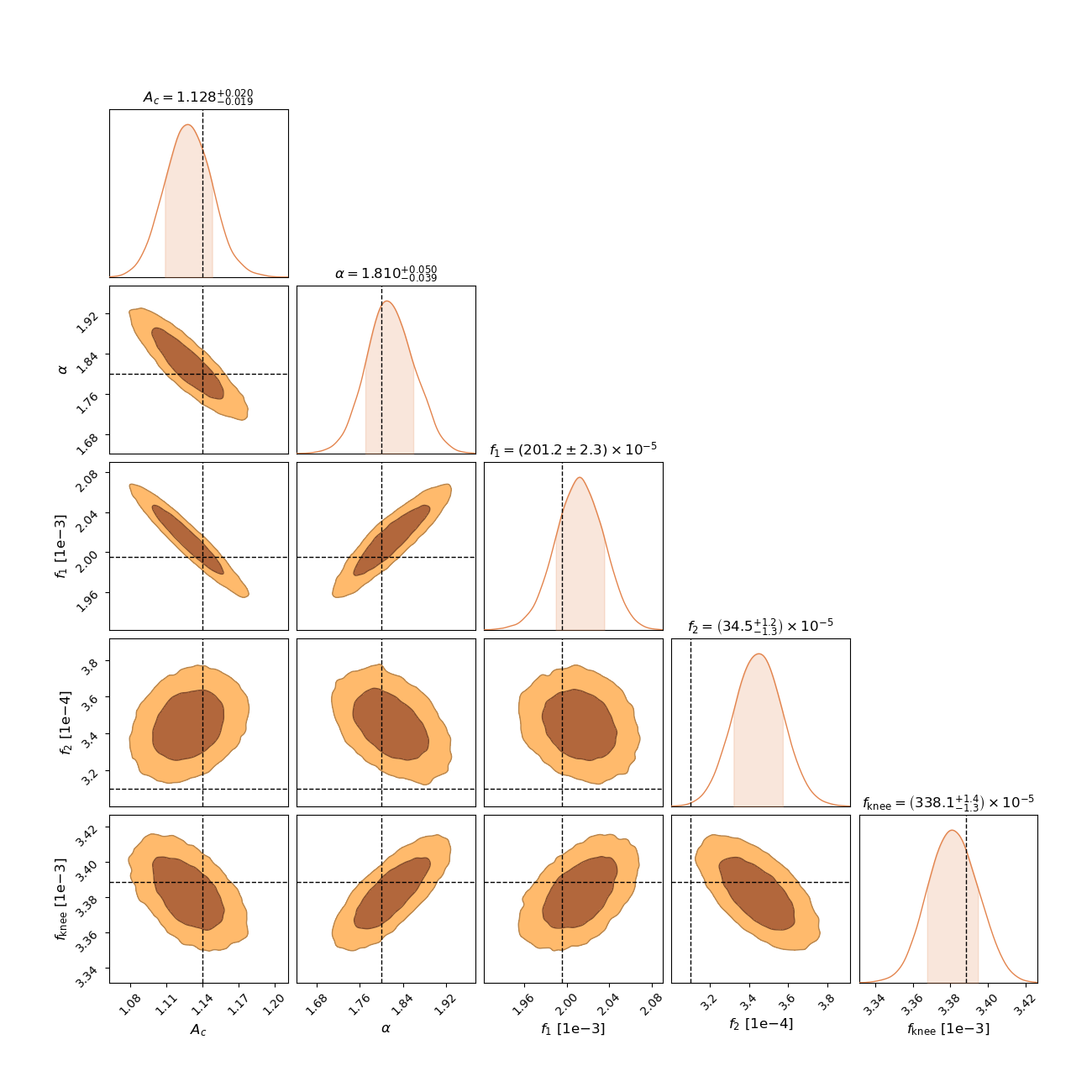}%
}
{
  \includegraphics[clip,width=0.9\columnwidth]{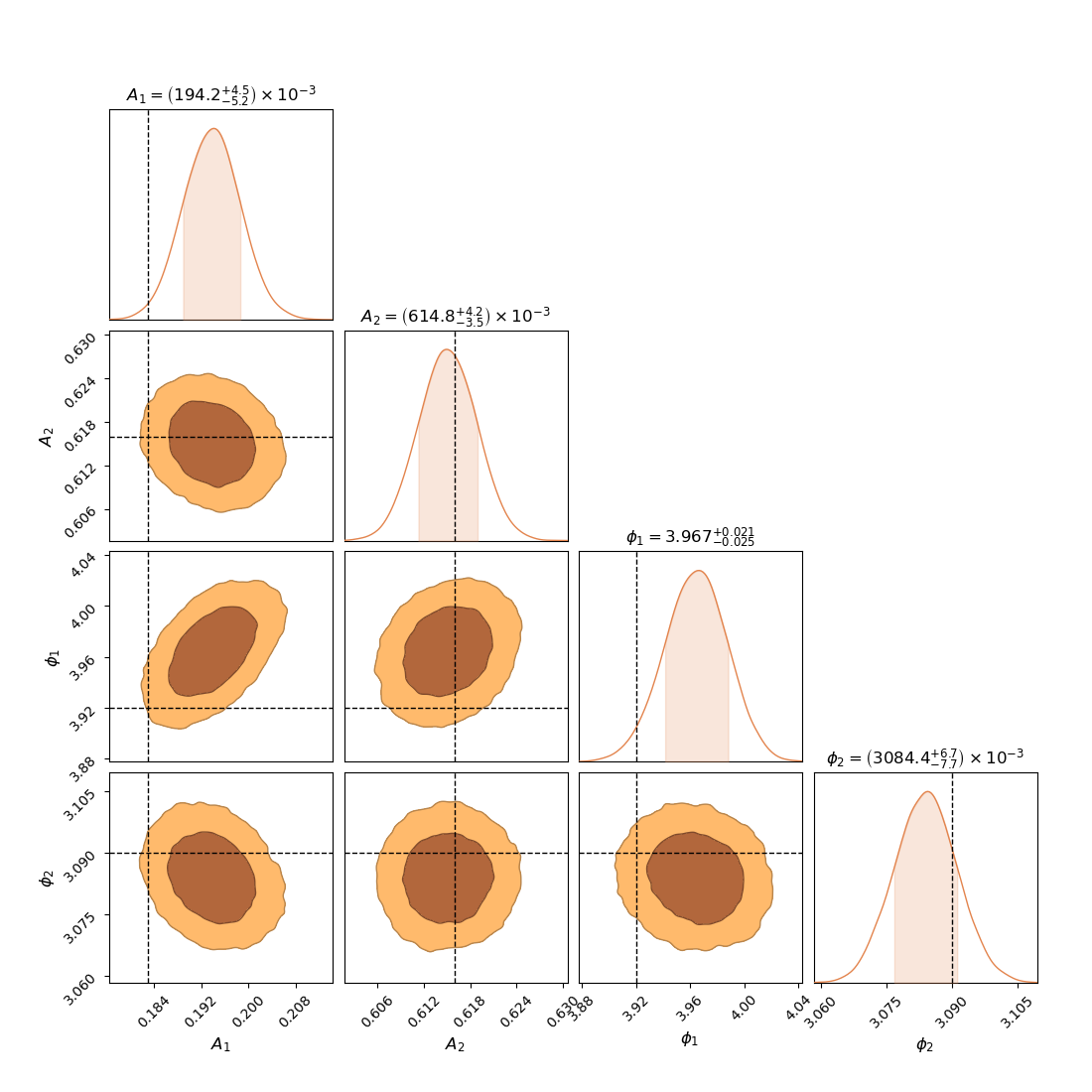}%
}

\caption{Example of recovered samples of the noise model posterior for augmented data, showing good parameter recovery. Parameters the top panel are the two amplitude parameters on the spectrum of the stationary instrument noise, the middle panel are the five parameters affecting the shape of the galactic foreground noise spectrum (see Fig.(\ref{fig: noise model f})), and the bottom panel are the amplitudes and phases in the time modulation function applied to the galactic foreground noise (see Fig.(\ref{fig:time mod})).}

\label{fig: noise corner}

\end{figure}

\begin{figure}
    \centering
    \includegraphics[width=1\linewidth]{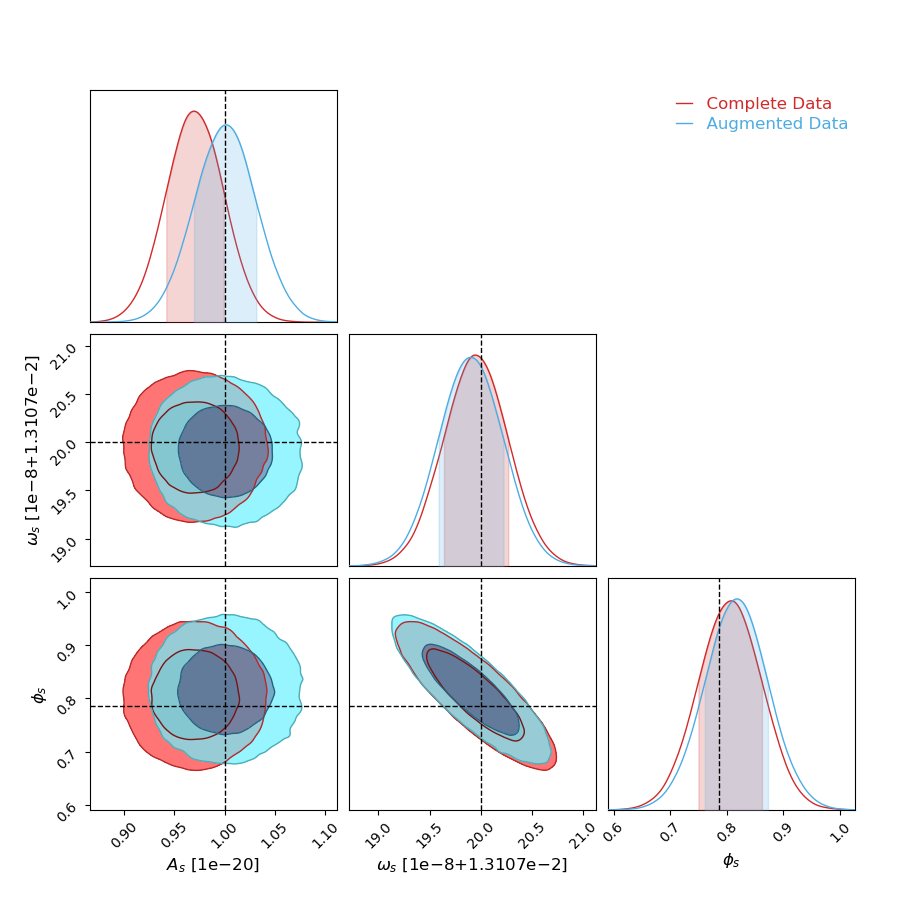}
    \caption{Example of recovered samples of the signal model (amplitude, frequency, and phase of a sinusoid) posterior for augmented data, showing good parameter recovery consistent with the complete data.}
    \label{fig: signal corner}
\end{figure}

\begin{figure}[htp]
{
  \includegraphics[clip,width=0.7\columnwidth]{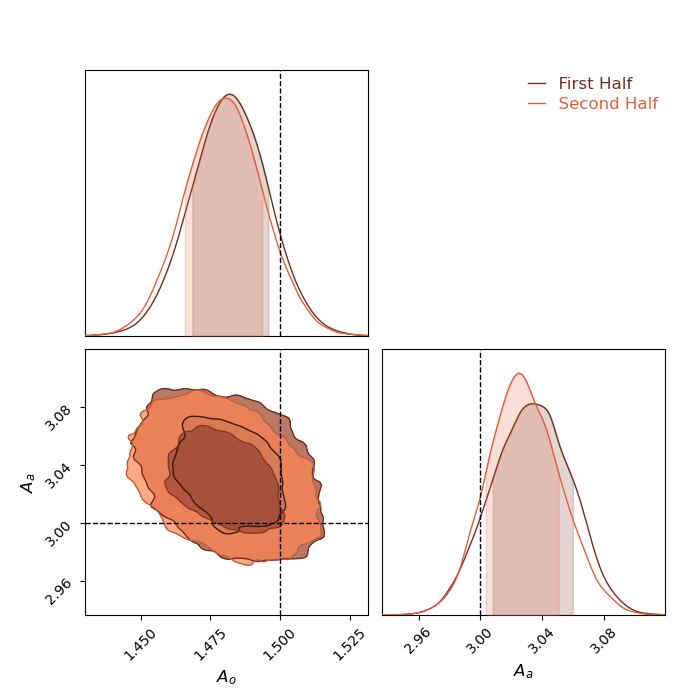}%
}

{
  \includegraphics[clip,width=0.9\columnwidth]{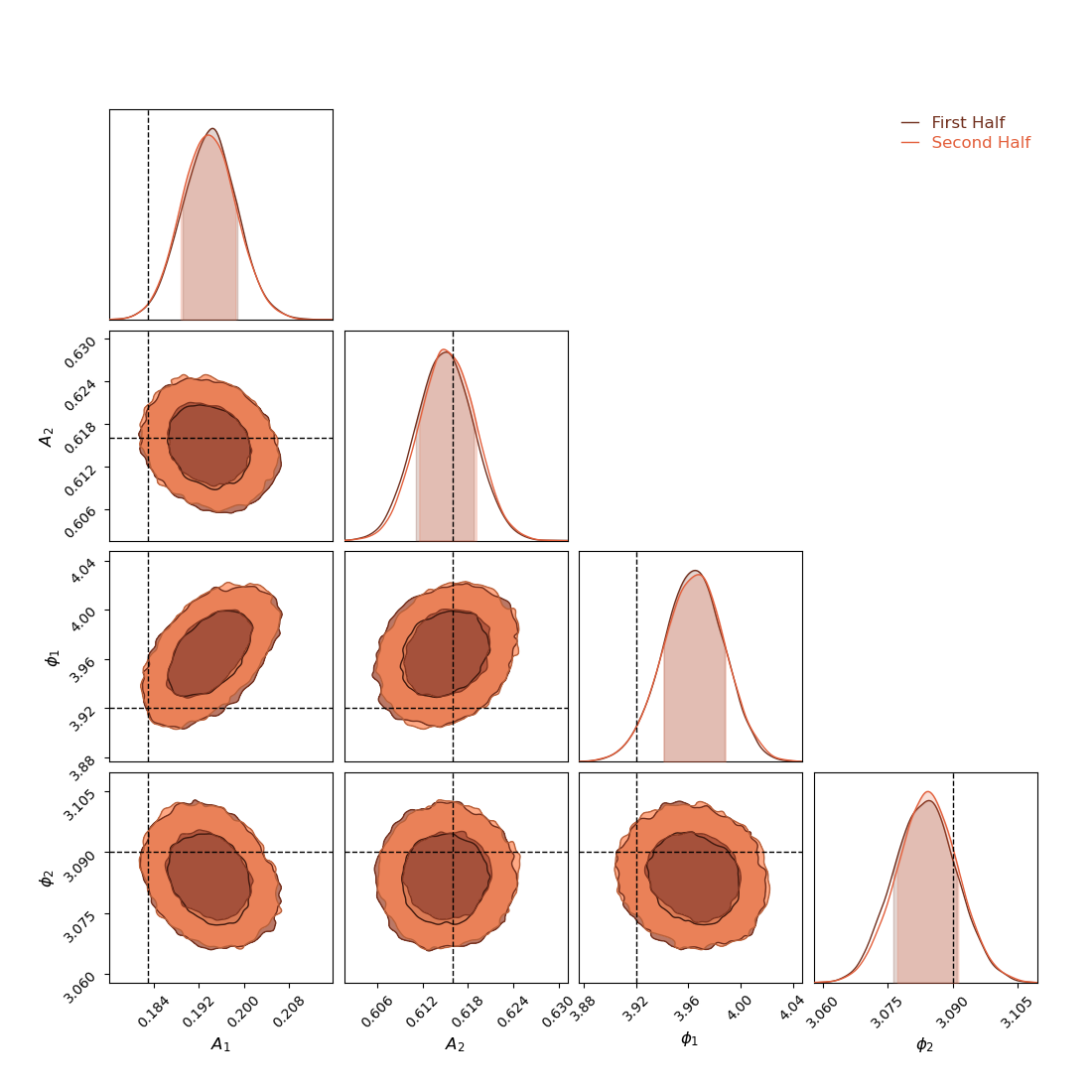}%
}
{
  \includegraphics[clip,width=0.9\columnwidth]{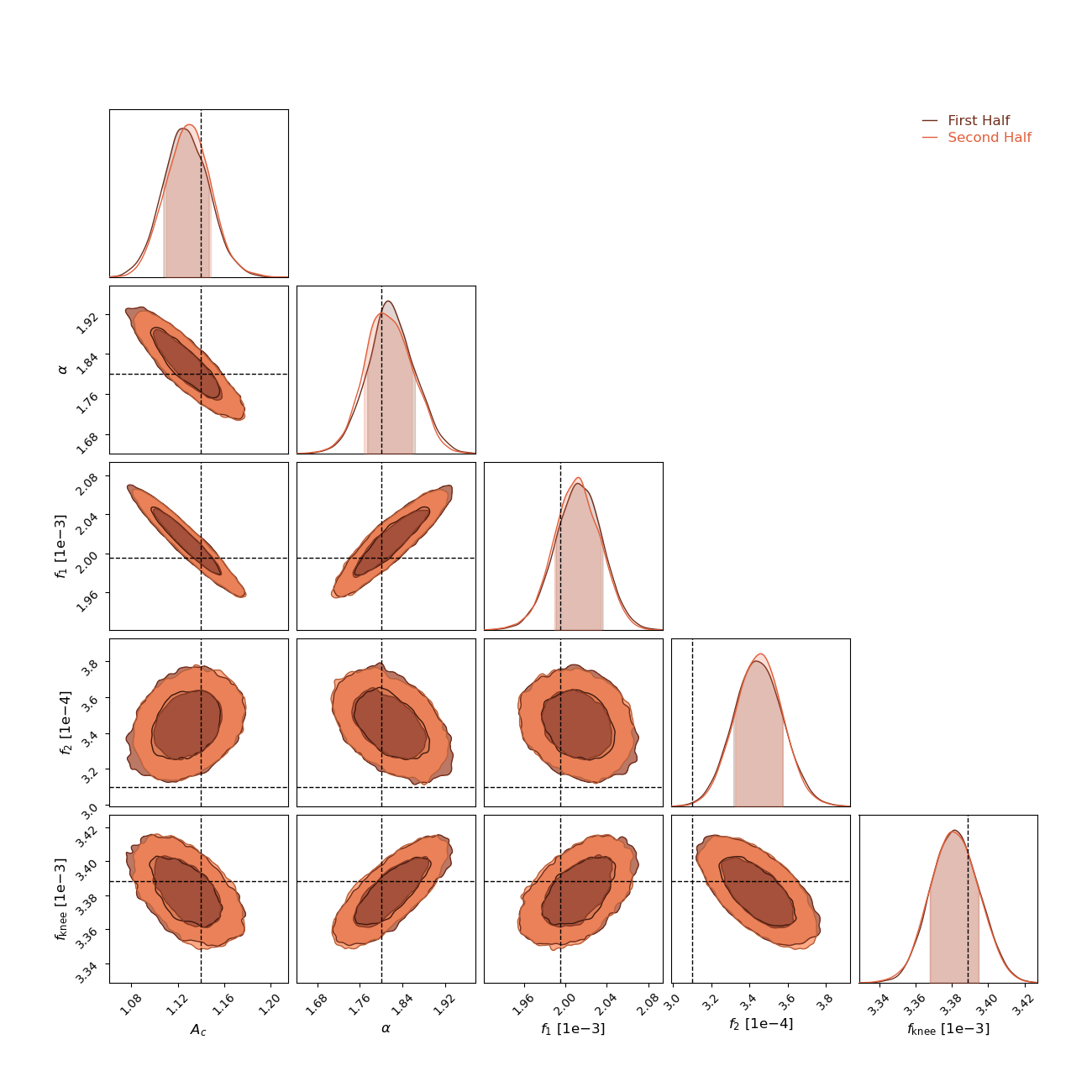}%
}

\caption{The first and second halves of the recovered noise model posterior samples displayed in Fig.(\ref{fig: noise corner}). Because the posteriors do not change significantly between the halves, we may conclude that the sampler has converged prior to $10^6$ iterations.}

\label{fig: noise corner stationary}

\end{figure}

We observe that the results of performing this treatment (see Fig.(\ref{fig: noise corner}), Fig.(\ref{fig: signal corner})) are consistent with the posterior samples from un-gapped data, with no discernible bias. We note that the sampler has converged by visually inspecting the parameter trace plots which show stationary exploration; this can also be observed qualitatively in the posteriors of the first and second halves of the samples being near-identical, see Fig.(\ref{fig: noise corner stationary}); the same story is seen in the signal posteriors. All signal parameters and noise hyper-parameters lie within the posterior distribution. The comparison is favorable with untreated data with no augmentation (which failed to converge, even initializing with the injected parameters). We also note broadening of the signal posteriors consistent with the loss of information due to the gaps; though this is not a significant effect, with roughly 10\% of the data missing for a monochrome sinusoid, the posteriors are only broadened by $\sim$5\% (see Appendix B). Posterior broadening may also occur in the case that more parameters with correlations are being sampled over. However, in this work no additional parameters are being added - even though we are ``sampling" over possible filling data, its only purpose is to remove spectral leakage. In the likelihood, the imputed signal is explicitly subtracted out, and the imputing noise is subtracted out via the noise covariance, i.e. ``subtraction equals division."~\cite{Lentati2013,Cornish2013,Romano2016}.

For the case of an evolving signal (e.g. an inspiraling black hole binary), the SNR is not evenly distributed. In this case, when gaps occur becomes relevant for the SNR accumulated since most of the SNR is concentrated in the merger period. Lost SNR alone is not a perfect proxy for determining uncertainty in this case, since the phase can be coherently traced through a gap, even at merger.

\subsection{Conclusions}


In this paper, we've introduced a new sampling technique that has made Bayesian data augmentation an efficient, accurate, and feasible solution. For the first time, to the best of our knowledge, we also applied augmentation in the context of a time-frequency domain analysis which is most capable at handling non-stationary processes. We demonstrated a prototype algorithm's success in a simulation of LISA-like Michelson-style data, accurately recovering the injected signal parameters and noise model hyper-parameters. Moreover, we've shown its use in eliminating spectral leakage at the edges of a dataset, and in handling noise state changes across a gap. 

This method is appealing for many reasons. There is no gap pattern for which this method ``fails" - increasing the amount of data lost to gaps only increases the uncertainty of the filling data. Parameter recovery will not be biased, only less certain. While this tool was developed primarily with the future LISA detector in mind, it is applicable to data from any future detector that encounters the joint issues of non-stationary noise and data gaps.

\section{Future Work}

In the future of time-frequency analysis, we will continue to explore modifications to the basis and their effects on data augmentation. We previously mentioned experimenting with the choice of the window function; e.g. choosing windows that are more localized in time might confer some advantages in augmentation, by virtue of the limited impact gaps would have in time (at the expense of less localization in frequency). 

As discussed, there exists techniques for on-source PSD estimation that will aid in initializing the state of the imputing data. In future implementations, we intend to integrate Bayesian data augmentation into a more advanced sampler that can robustly initialize and burn-in to the locus of the \textit{maximum a posteriori}. We also intend to explore alternative ``small-jump" fill proposals to use in conjunction with the conditional distribution.

In this prototype implementation, we considered a single channel of Michelson-style data. For the actual LISA, TDI will be favorable over traditional Michelson-style channels because of its suppression of laser phase noise. This means that the behavior of gaps will be slightly different (as discussed in Sec.(\ref{sec: data gaps})), and the general spectral character of the noise will be blue instead of red. We don't believe significant modification to this procedure of gap filling is necessary to fully implement it in the analysis of TDI variables. If anything, it may make the initialization easier in that the filling data will be more locally correlated.

Ultimate global fit pipelines to be used on the real LISA data will need to contend with gaps. Currently, the Global LISA Analysis Software Suite (\texttt{GLASS}) which has successfully analyzed LISA data challenge \texttt{LDC2a-v2}~\cite{LDC} has not run on LISA data challenges with gaps; we intend to integrate this tool into the software.

\section{Acknowledgments}

We would like to thank Guido Mueller and Martin Hewitson for their discussion of the point-ahead angle mechanism. This work was supported under NASA LISA Preparatory Science Grant 80NSSC19K0320. 

\section*{Appendix A \\ The WDM Wavelet Transformation Matrix}
\label{sec:appA}


As mentioned in Sec.(\ref{sec: tf analysis}), since the mappings use the WDM wavelet matrix, rather than the FFT based transforms developed in  previous works, we will briefly describe the construction and properties of the matrix.

\subsection{Matrix Construction}

As in Fourier analysis, we may construct a transformation matrix, $\bm{W}_{N \times N}(\mathbb{R})$ which maps the data time series $\bm{d}$ to wavelet amplitudes $\bm{a}$,

\begin{equation}
    \bm{a} = \bm{W\,d}\,.
\end{equation}

Note that this transformation matrix is square, so it maps the time series to an $N$-vector rather than to the 2D $N_f \times N_t$ pixel grid. This transformed $N$-vector $\bm{a} = \{a_J\}$ may be mapped to the pixel grid ($J \leftrightarrow (n,m)$) in the following way, according to this work's configuration of the wavelet matrix,

\begin{equation}
    J = 
    \begin{cases}
        n N_f + m, & m \in [1,N_f-1] \\
        n N_f, & m = 0 \,,\, n \in [0,N_t/2-1]\\
        & m = N_f \,,\, n \in [N_t/2,N_t-1]
    \end{cases}
\end{equation}

\noindent where `$J$' is a condensed index of the time-frequency domain. \textit{Nota bene} that for the extremal frequency layers $m = 0,N_f$ there's only $N_t/2$ wavelets each since in those layers the wavelets cover twice the temporal extent and half the frequency bandwidth as the ``core" frequency layers $m\in [1,N_f-1]$, thus we still have $N_f\times N_t = N$. In practice, it seems easiest to make an array of size $(N_f+1)\times N_t$ to store the amplitudes; where the $m=0$ amplitudes are stored in the first $N_t/2$ pixels of the $m=0$ frequency layer, and the $m=N_f$ are stored in the last $N_t/2$ pixels of the $m=N_f+1$ layer. In this work we however neglect the extremal frequencies.

The matrix that relates the time and wavelet domains is constructed using wavelets in the time domain. The transform between frequency and wavelet domains would require wavelets in the frequency domain, but here we only describe the relationship relevant to this work (though the computation is similar). In the time domain, the WDM wavelets have the following form~\cite{Necula12},

\begin{widetext}
\begin{equation}
    \psi_{nm} (t) =
    \begin{cases}
        \phi(t-2n\Delta T)\,, & m = 0 \\
        \sqrt{2}(-1)^{nm} \,\mathrm{cos}\,(\pi m t /\Delta T) \phi(t-n\Delta T)\,, & m+n = \mathrm{even} \\
        \sqrt{2}\,\mathrm{sin}\,(\pi m t /\Delta T) \phi(t-n\Delta T)\,, & m+n = \mathrm{odd} \\
        \mathrm{cos}\,\left[\pi(t-2n\Delta T)\right]\phi(t-2n\Delta T)\,, & m = N_f
        
    \end{cases}
\end{equation}
\end{widetext}

\noindent where $\phi = \phi(t)$ is the Meyer window function in the time domain, centered at $t = 0$. It's most naturally defined in the frequency domain ($\omega = 2\pi f$),

\begin{equation}
    \Phi(\omega) = 
    \begin{cases}
        \frac{1}{\sqrt{\Delta\Omega}}\,, &|\omega| < A \\
        \frac{1}{\sqrt{\Delta\Omega}}\,\mathrm{cos}\,\left[\frac{\pi}{2}\nu_d\left(\frac{|\omega|-A}{B}\right)\right]\,, A\leq &|\omega|\leq A+B \\
    \end{cases}
    \,,
\end{equation}

\noindent where $A,B$ are constrained by $2A+B = \Delta\Omega = 2\pi\Delta F$. We choose $A=0$ and $B=\Delta\Omega$. $\nu_d(x)$ is the normalized incomplete Beta function,

\begin{equation}
    \nu_d(x) = \frac{\int_0^xy^{d-1}(1-y)^{d-1} \dd y}{\int_0^1y^{d-1}(1-y)^{d-1} \dd y}\,,
\end{equation}

\noindent we use $d=6$. With these choices for the window parameters, we can choose $q=8$ without incurring significant errors in truncation. The morphology of the wavelets, and thus the analysis, is not strongly sensitive to the choice of their shape parameters, $A,B,d$. What's more relevant is the choice of time-frequency resolution, $\Delta T,\Delta F$, since it affects the locally-stationary approximation, and sets the wavelet kernel size.

To motivate the structure of the wavelet matrix, first consider the usual way of computing transformation coefficients, with the inner-product between the basis functions and the data,

\begin{equation}
    a_{nm} = \sum_{i = 0}^{N-1} \psi_{nm}(t_i)\,d(t_i)\,.
    \label{eq: anm}
\end{equation}

For every time pixel, the transform needs to cover the whole frequency range, so the substructure of the matrix includes $N_t$ blocks of size $N_f \times K$. Since the $m = 0,N_f$ frequency layers have pixels with twice the temporal extent, only $N_t/2$ wavelets are needed for each of the edge frequency layers. We choose to place the $m = 0$ wavelets in the first $N_t/2$ blocks and $m = N_f$ in the last half. Similar to a discrete Fourier transform, we require that the wavelets are periodically extended at the boundaries.

The indicial form of the WDM transformation matrix is,

\begin{equation}
    W_{JI} = \psi_{nm}\left[t_{I}\right]\times
    \begin{cases}
        \Pi\left[\frac{1}{K}\left(t_I-2J+\frac{N}{2}\right)\right]\,, & m = 0 \\
        \Pi\left[\frac{1}{K}\left(t_I-2J-\frac{N}{2}\right)\right]\,, & m = N_f \\
        \Pi\left[\frac{1}{K}\left(t_I-J+\frac{N}{2}\right)\right]\,, & m \in [1,N_f-1] \\
    \end{cases}
\end{equation}

\noindent where $J \rightarrow (n,m)$ the time-frequency indices as mapped from earlier, and $\Pi$ is the $N$-periodically-wrapped rectangular function, of width $K$, centered on $N/2$. The wavelets $\psi_{nm}$ also periodically wrap. For $K = N$, $\Pi = 1,\,\forall t$.

\begin{figure}
    \centering
    \includegraphics[width=1\linewidth]{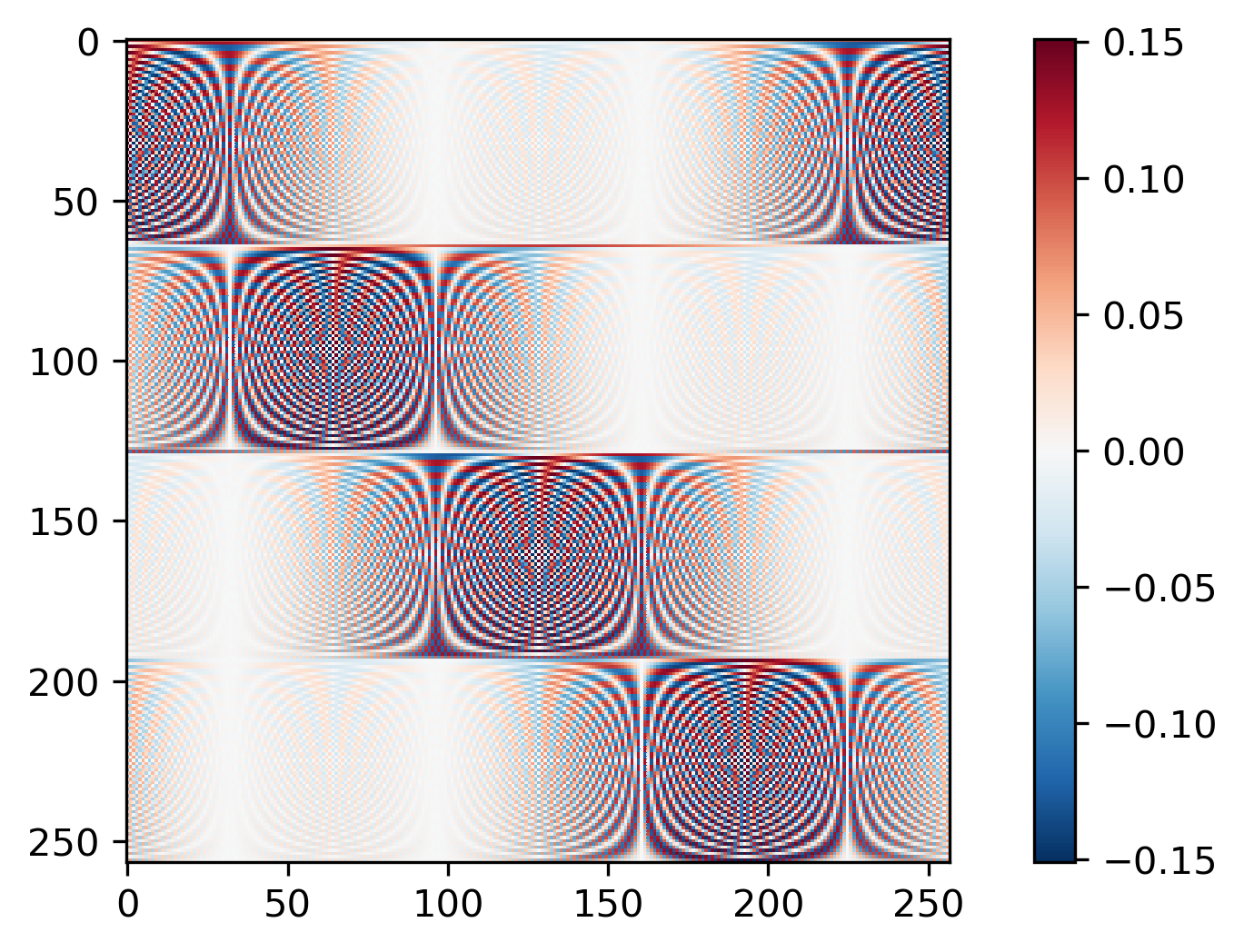}
    \caption{Plotted element values for the WDM wavelet transformation matrix $\bm{W}$, with $N_f=64,\,N_t=4\,,q=2$ - to illustrate its structure.}
    \label{fig:wmatrix}
\end{figure}

Because the form of the WDM wavelets is invariant under translations in the central time, the transformation matrix exhibits a repeating sub-structure, see Fig.(\ref{fig:wmatrix}). This means that rather than compute and store an $N\times N$ matrix, one need only compute two of the sub-structure blocks - corresponding to the two $n+m = $ even/odd wavelet combinations for all $m$ - and then map the indices to perform the transformation operation or obtain a subset of the W-matrix, as needed in the neighborhood approximation to the time domain covariance.

\subsection{Matrix Properties}

The transformation matrix has the useful property of being orthogonal ($\bm{W}^\mathrm{T} = \bm{W}^{-1}$) which is derived from the orthogonality of the wavelets,

\begin{equation}
    \bm{WW}^\mathrm{T} = \mathbb{I}_N\,.
\end{equation}

\noindent In practice, the wavelets are truncated in time, the orthogonality will not be perfect, but the departure from perfect orthogonality can be kept small using suitable choices for the wavelet parameters. Being a unitary transformation, the wavelet matrix has unit determinant,

\begin{equation}
    \mathrm{det}(\bm{W}) = \pm 1\,.
\end{equation}

To transform a 2D object into the wavelet domain (such as the covariance matrix), it is multiplied with the wavelet matrix and its inverse (which is more cheaply computed using the orthogonal property),

\begin{equation}
    \bm{\Sigma} = \bm{W}^{-1}\bm{\Lambda\,W} = \bm{W}^\mathrm{T}\bm{\Lambda\,W}\,,
\end{equation}

\noindent where we have condensed the wavelet domain covariance matrix $\Lambda_{(nm)(n'm')}\rightarrow\Lambda_{JI}$ to make it $N\times N$. And its inverse,

\begin{equation}
    \bm{\Sigma}^{-1} = \bm{W}^\mathrm{T}\bm{\Lambda}^{-1}\bm{W}\,.
\end{equation}

\noindent Because $\bm{\Lambda}$ is diagonal, this makes computing the inverse covariance matrix in the time domain cost only $N^2$.

\section*{Appendix B \\ Calculation of SNR}

To assuage concerns that augmenting data will affect the SNR, we provide results demonstrating that in our application the SNR of the observed data is preserved and give an explicit definition. Formally in the wavelet-domain the complete SNR is,

\begin{equation}
     \mathrm{SNR}^2 \equiv (\bm{h}|\bm{h}) = \sum_{n,m} \frac{h_{nm}^2}{S(f_m,t_n)}\,.
\end{equation}

\noindent However, we can only accumulate SNR for times that are observed. In the wavelet domain, only fractions of the wavelet pixel temporal span would be observed. Therefore, the SNR should be calculated in the time domain,

\begin{equation}\label{snrgap}
    \mathrm{SNR}^2 = \sum_{t_i\in T_o}(h^w(t_i))^2 = \bm{h}_o^\mathrm{T}(t)\bm{\Sigma}_{oo}^{-1}\bm{h}_o(t)
\end{equation}

\noindent where $T_o= \{t_o\}$ is the set of observed time indices, and $\bm{h}^w(t)$ is the whitened signal in the time domain. In practice this method of computing the SNR is expensive as, $\bm{\Sigma}_{oo}$ costs $\mathcal{O}(N^3)$ to invert (assuming $N_o \sim N$). Instead, the signal may be whitened in the wavelet domain, then transformed to the time domain at cost $\mathcal{O}(N\mathrm{log}N)$ using the fast transform,
\begin{equation}
    h_{nm} \xrightarrow{\mathrm{whiten}} h^w_{nm} = \frac{h_{nm}} {\sqrt{S(f_m,t_n)}} \xrightarrow{\mathrm{transform}} \bm{h}^w(t)\, .
\end{equation}

Because of the loss of information that gaps represent, we should see a broadening of the posterior distribution that roughly corresponds with the loss of SNR (especially for non-chirping signals with a largely uniform distribution of power in their time-frequency extent). In our testing data, we should see a broadening factor consistent with the ratio of SNR's, $\mathrm{SNR_{complete}}/\mathrm{SNR_{gapped}}$.

\bibliographystyle{utphys}
\bibliography{refs}

@article{LISA:2024hlh,
    author = "Colpi, Monica and others",
    collaboration = "LISA",
    title = "{LISA Definition Study Report}",
    eprint = "2402.07571",
    archivePrefix = "arXiv",
    primaryClass = "astro-ph.CO",
    month = "2",
    year = "2024"
}

@ARTICLE{lisa_prop,
       author = {{Amaro-Seoane}, Pau and {Audley}, Heather and {Babak}, Stanislav and others},
        title = "{Laser Interferometer Space Antenna}",
      journal = {arXiv e-prints},
         year = 2017,
        month = feb,
          eid = {arXiv:1702.00786},
        pages = {arXiv:1702.00786},
          doi = {10.48550/arXiv.1702.00786},
archivePrefix = {arXiv},
       eprint = {1702.00786},
 primaryClass = {astro-ph.IM},
       adsurl = {https://ui.adsabs.harvard.edu/abs/2017arXiv170200786A}
}

@article{pathfinder,
  title = {Beyond the Required LISA Free-Fall Performance: New LISA Pathfinder Results down to $20\text{ }\text{ }\ensuremath{\mu}\mathrm{Hz}$},
  author = {Armano, M. and Audley, H. and Baird, J. and others},
  journal = {Phys. Rev. Lett.},
  volume = {120},
  issue = {6},
  pages = {061101},
  numpages = {7},
  year = {2018},
  month = {Feb},
  publisher = {American Physical Society},
  doi = {10.1103/PhysRevLett.120.061101},
  url = {https://link.aps.org/doi/10.1103/PhysRevLett.120.061101}
}

@article{Cornish_2020,
	doi = {10.1103/physrevd.102.124038},
  
	url = {https://doi.org/10.1103%2Fphysrevd.102.124038},
  
	year = 2020,
	month = {dec},
  
	publisher = {American Physical Society ({APS})},
  
	volume = {102},
  
	number = {12},
  
	author = {Neil J. Cornish},
  
	title = {Time-frequency analysis of gravitational wave data},
  
	journal = {Physical Review D}
}

@article{Necula12,
    author = "Necula, V. and Klimenko, S. and Mitselmakher, G.",
    editor = "Hannam, Mark and Sutton, Patrick and Hild, Stefan and van den Broeck, Chris",
    title = "{Transient analysis with fast Wilson-Daubechies time-frequency transform}",
    doi = "10.1088/1742-6596/363/1/012032",
    journal = "J. Phys. Conf. Ser.",
    volume = "363",
    pages = "012032",
    year = "2012"
}

@article{lisa_white,
	doi = {https://doi.org/10.1007/s41114-022-00041-y},
  
	url = {https://doi.org/10.1007/s41114-022-00041-y},
  
	year = 2023,
	month = {mar},
  
	publisher = {Springer Link},
  
	volume = {26},
  
	number = {2},
  
	author = {Amaro-Seoane, P. and Andrews, J. and Arca Sedda, M. and others},
  
	title = {Astrophysics with the Laser Interferometer Space Antenna},
  
	journal = {Living Rev Relativ}
}

@ARTICLE{lisa22,
       author = {{Bayle}, Jean-Baptiste and {Bonga}, B{\'e}atrice and {Caprini}, Chiara and {Doneva}, Daniela and {Muratore}, Martina and {Petiteau}, Antoine and {Rossi}, Elena and {Shao}, Lijing},
        title = "{Overview and progress on the Laser Interferometer Space Antenna mission}",
      journal = {Nature Astronomy},
         year = 2022,
        month = dec,
       volume = {6},
        pages = {1334-1338},
          doi = {10.1038/s41550-022-01847-0},
       adsurl = {https://ui.adsabs.harvard.edu/abs/2022NatAs...6.1334B}
}

@article{Baghi_2019,
	doi = {10.1103/physrevd.100.022003},
  
	url = {https://doi.org/10.1103%2Fphysrevd.100.022003},
  
	year = 2019,
	month = {jul},
  
	publisher = {American Physical Society ({APS})},
  
	volume = {100},
  
	number = {2},
  
	author = {Quentin Baghi and James Ira Thorpe and Jacob Slutsky and John Baker and Tito Dal Canton and Natalia Korsakova and Nikos Karnesis},
  
	title = {Gravitational-wave parameter estimation with gaps in {LISA}: A Bayesian data augmentation method},
  
	journal = {Physical Review D}
}

@article{Blelly_2021,
	doi = {10.1093/mnras/stab3314},
  
	url = {https://doi.org/10.1093%2Fmnras%2Fstab3314},
  
	year = 2021,
	month = {nov},
  
	publisher = {Oxford University Press ({OUP})},
  
	volume = {509},
  
	number = {4},
  
	pages = {5902--5917},
  
	author = {Aurore Blelly and J{\'{e}
}r{\^{o}}me Bobin and Herv{\'{e}} Moutarde},
  
	title = {Sparse data inpainting for the recovery of Galactic-binary gravitational wave signals from gapped data},
  
	journal = {Monthly Notices of the Royal Astronomical Society}
}

@article{Capano:2021,
    author = "Capano, Collin D. and Cabero, Miriam and Westerweck, Julian and Abedi, Jahed and Kastha, Shilpa and Nitz, Alexander H. and Wang, Yi-Fan and Nielsen, Alex B. and Krishnan, Badri",
    title = "{Multimode Quasinormal Spectrum from a Perturbed Black Hole}",
    eprint = "2105.05238",
    archivePrefix = "arXiv",
    primaryClass = "gr-qc",
    doi = "10.1103/PhysRevLett.131.221402",
    journal = "Phys. Rev. Lett.",
    volume = "131",
    number = "22",
    pages = "221402",
    year = "2023"
}

@article{zackay2021,
  title = {Detecting gravitational waves in data with non-stationary and non-Gaussian noise},
  author = {Zackay, Barak and Venumadhav, Tejaswi and Roulet, Javier and Dai, Liang and Zaldarriaga, Matias},
  journal = {Phys. Rev. D},
  volume = {104},
  issue = {6},
  pages = {063034},
  numpages = {17},
  year = {2021},
  month = {Sep},
  publisher = {American Physical Society},
  doi = {10.1103/PhysRevD.104.063034},
  url = {https://link.aps.org/doi/10.1103/PhysRevD.104.063034}
}

@article{Venumadhav2019,
    author = "Venumadhav, Tejaswi and Zackay, Barak and Roulet, Javier and Dai, Liang and Zaldarriaga, Matias",
    title = "{New search pipeline for compact binary mergers: Results for binary black holes in the first observing run of Advanced LIGO}",
    eprint = "1902.10341",
    archivePrefix = "arXiv",
    primaryClass = "astro-ph.IM",
    doi = "10.1103/PhysRevD.100.023011",
    journal = "Phys. Rev. D",
    volume = "100",
    number = "2",
    pages = "023011",
    year = "2019"
}

@ARTICLE{Isi2021,
       author = {{Isi}, Maximiliano and {Farr}, Will M.},
        title = "{Analyzing black-hole ringdowns}",
      journal = {arXiv e-prints},
         year = 2021,
        month = jul,
          eid = {arXiv:2107.05609},
        pages = {arXiv:2107.05609},
          doi = {10.48550/arXiv.2107.05609},
archivePrefix = {arXiv},
       eprint = {2107.05609},
 primaryClass = {gr-qc},
       adsurl = {https://ui.adsabs.harvard.edu/abs/2021arXiv210705609I}
}

@ARTICLE{wang24,
       author = {{Wang}, Lu and {Chen}, Hong-Yu and {Lyu}, Xiangyu and {Li}, En-Kun and {Hu}, Yi-Ming},
        title = "{Window and inpainting: dealing with data gaps for TianQin}",
      journal = {arXiv e-prints},
         year = 2024,
        month = may,
          eid = {arXiv:2405.14274},
        pages = {arXiv:2405.14274},
          doi = {10.48550/arXiv.2405.14274},
archivePrefix = {arXiv},
       eprint = {2405.14274},
 primaryClass = {gr-qc},
       adsurl = {https://ui.adsabs.harvard.edu/abs/2024arXiv240514274W}
}

@article{Mao2024,
    author = "Mao, Ruiting and Lee, Jeong Eun and Edwards, Matthew C.",
    title = "{Novel stacked hybrid autoencoder for imputing LISA data gaps}",
    eprint = "2410.05571",
    archivePrefix = "arXiv",
    primaryClass = "gr-qc",
    doi = "10.1103/PhysRevD.111.024067",
    journal = "Phys. Rev. D",
    volume = "111",
    number = "2",
    pages = "024067",
    year = "2025"
}

@book{eaton1983,
  title={Multivariate Statistics: A Vector Space Approach},
  author={Eaton, M.L.},
  isbn={9780471027768},
  lccn={83001215},
  series={Probability and Statistics Series},
  url={https://books.google.com/books?id=1CvvAAAAMAAJ},
  year={1983},
  publisher={Wiley}
}

@article{Robson2018,
    author = "Robson, Travis and Cornish, Neil J. and Liu, Chang",
    title = "{The construction and use of LISA sensitivity curves}",
    eprint = "1803.01944",
    archivePrefix = "arXiv",
    primaryClass = "astro-ph.HE",
    doi = "10.1088/1361-6382/ab1101",
    journal = "Class. Quant. Grav.",
    volume = "36",
    number = "10",
    pages = "105011",
    year = "2019"
}

@article{Digman2022,
    author = "Digman, Matthew C. and Cornish, Neil J.",
    title = "{LISA Gravitational Wave Sources in a Time-varying Galactic Stochastic Background}",
    eprint = "2206.14813",
    archivePrefix = "arXiv",
    primaryClass = "astro-ph.IM",
    doi = "10.3847/1538-4357/ac9139",
    journal = "Astrophys. J.",
    volume = "940",
    number = "1",
    pages = "10",
    year = "2022"
}

@article{dey2021,
  title = {Effect of data gaps on the detectability and parameter estimation of massive black hole binaries with LISA},
  author = {Dey, Kallol and Karnesis, Nikolaos and Toubiana, Alexandre and Barausse, Enrico and Korsakova, Natalia and Baghi, Quentin and Basak, Soumen},
  journal = {Phys. Rev. D},
  volume = {104},
  issue = {4},
  pages = {044035},
  numpages = {25},
  year = {2021},
  month = {Aug},
  publisher = {American Physical Society},
  doi = {10.1103/PhysRevD.104.044035},
  url = {https://link.aps.org/doi/10.1103/PhysRevD.104.044035}
}

@article{Ricciardone2016,
    author = "Ricciardone, Angelo",
    editor = "Giardini, Domencio and Jetzer, Philippe",
    title = "{Primordial Gravitational Waves with LISA}",
    eprint = "1612.06799",
    archivePrefix = "arXiv",
    primaryClass = "astro-ph.CO",
    doi = "10.1088/1742-6596/840/1/012030",
    journal = "J. Phys. Conf. Ser.",
    volume = "840",
    number = "1",
    pages = "012030",
    year = "2017"
}

@article{Cornish2006,
    author = "Cornish, Neil J. and Porter, Edward K.",
    title = "{The Search for supermassive black hole binaries with LISA}",
    eprint = "gr-qc/0612091",
    archivePrefix = "arXiv",
    doi = "10.1088/0264-9381/24/23/001",
    journal = "Class. Quant. Grav.",
    volume = "24",
    pages = "5729--5755",
    year = "2007"
}

@article{Burke2025a,
    author = "Burke, Ollie and Marsat, Sylvain and Gair, Jonathan R. and Katz, Michael L.",
    title = "{Mind the gap: addressing data gaps and assessing noise mismodeling in LISA}",
    eprint = "2502.17426",
    archivePrefix = "arXiv",
    primaryClass = "gr-qc",
    month = "2",
    year = "2025"
}

@article{Karnesis2021,
    author = "Karnesis, Nikolaos and Babak, Stanislav and Pieroni, Mauro and Cornish, Neil and Littenberg, Tyson",
    title = "{Characterization of the stochastic signal originating from compact binary populations as measured by LISA}",
    eprint = "2103.14598",
    archivePrefix = "arXiv",
    primaryClass = "astro-ph.IM",
    doi = "10.1103/PhysRevD.104.043019",
    journal = "Phys. Rev. D",
    volume = "104",
    number = "4",
    pages = "043019",
    year = "2021"
}

@article{Houba2024,
    author = "Houba, Niklas and Bayle, Jean-Baptiste and Vallisneri, Michele",
    title = "{Robust Bayesian inference with gapped LISA data using all-in-one TDI-$\infty$}",
    eprint = "2412.20793",
    archivePrefix = "arXiv",
    primaryClass = "astro-ph.IM",
    month = "12",
    year = "2024"
}

@article{Castelli2024,
    author = "Castelli, Eleonora and Baghi, Quentin and Baker, John G. and Slutsky, Jacob and Bobin, J\'er\^ome and Karnesis, Nikolaos and Petiteau, Antoine and Sauter, Orion and Wass, Peter and Weber, William J.",
    title = "{Extracting gravitational wave signals from LISA data in the presence of artifacts}",
    eprint = "2411.13402",
    archivePrefix = "arXiv",
    primaryClass = "gr-qc",
    doi = "10.1088/1361-6382/adb931",
    journal = "Class. Quant. Grav.",
    volume = "42",
    number = "6",
    pages = "065018",
    year = "2025"
}

@article{LISA2024,
    author = "Colpi, Monica and others",
    collaboration = "LISA",
    title = "{LISA Definition Study Report}",
    eprint = "2402.07571",
    archivePrefix = "arXiv",
    primaryClass = "astro-ph.CO",
    month = "2",
    year = "2024"
}

@article{Cornish2014,
    author = "Cornish, Neil J. and Littenberg, Tyson B.",
    title = "{BayesWave: Bayesian Inference for Gravitational Wave Bursts and Instrument Glitches}",
    eprint = "1410.3835",
    archivePrefix = "arXiv",
    primaryClass = "gr-qc",
    doi = "10.1088/0264-9381/32/13/135012",
    journal = "Class. Quant. Grav.",
    volume = "32",
    number = "13",
    pages = "135012",
    year = "2015"
}

@article{Littenberg2023,
    author = "Littenberg, Tyson B. and Cornish, Neil J.",
    title = "{Prototype global analysis of LISA data with multiple source types}",
    eprint = "2301.03673",
    archivePrefix = "arXiv",
    primaryClass = "gr-qc",
    doi = "10.1103/PhysRevD.107.063004",
    journal = "Phys. Rev. D",
    volume = "107",
    number = "6",
    pages = "063004",
    year = "2023"
}

@book{Stark1994,
author = {Stark, Henry and Woods, John},
year = {1994},
month = {01},
pages = {},
title = {Probability, Random Processes, and Estimation Theory for Engineers},
volume = {90},
journal = {Englewood Cliffs: Prentice Hall, 1986},
doi = {10.2307/2291115}
}

@article{LDC, 
    author={Le Jeune, Maude and Babak, Stanislav},
    title="{LISA Data Challenge Sangria (LDC2a)}", 
    url = "doi.org/10.5281/zenodo.7132178"
}

@article{Mallat98,
author = {St{\'e}phane Mallat and George Papanicolaou and Zhifeng Zhang},
title = {{Adaptive covariance estimation of locally stationary processes}},
volume = {26},
journal = {The Annals of Statistics},
number = {1},
publisher = {Institute of Mathematical Statistics},
pages = {1 -- 47},
year = {1998},
doi = {10.1214/aos/1030563977},
URL = {https://doi.org/10.1214/aos/1030563977}
}

@article{Muratore2025,
    author = "Muratore, Martina and Gair, Jonathan and Hartwig, Olaf and Katz, Michael L. and Toubiana, Alexandre",
    title = "{Pipeline for searching and fitting instrumental glitches in LISA data}",
    eprint = "2505.19870",
    archivePrefix = "arXiv",
    primaryClass = "gr-qc",
    doi = "10.1103/1sj2-219n",
    journal = "Phys. Rev. D",
    volume = "112",
    number = "6",
    pages = "063041",
    year = "2025"
}

@article{Burke2025b,
    author = "Burke, Ollie and Muratore, Martina and Woan, Graham",
    title = "{The impact of missing data on the construction of LISA Time Delay Interferometry Michelson variables}",
    eprint = "2510.06406",
    archivePrefix = "arXiv",
    primaryClass = "gr-qc",
    month = "10",
    year = "2025"
}

@ARTICLE{Lentati2013,
       author = {{Lentati}, Lindley and {Alexander}, P. and {Hobson}, M.~P. and {Taylor}, S. and {Gair}, J. and {Balan}, S.~T. and {van Haasteren}, R.},
        title = "{Hyper-efficient model-independent Bayesian method for the analysis of pulsar timing data}",
      journal = {\prd},
         year = 2013,
        month = may,
       volume = {87},
       number = {10},
          eid = {104021},
        pages = {104021},
          doi = {10.1103/PhysRevD.87.104021},
archivePrefix = {arXiv},
       eprint = {1210.3578},
 primaryClass = {astro-ph.IM},
       adsurl = {https://ui.adsabs.harvard.edu/abs/2013PhRvD..87j4021L}
}

@article{Cornish2013,
    author = "Cornish, Neil J. and Romano, Joseph D.",
    title = "{Towards a unified treatment of gravitational-wave data analysis}",
    eprint = "1305.2934",
    archivePrefix = "arXiv",
    primaryClass = "gr-qc",
    doi = "10.1103/PhysRevD.87.122003",
    journal = "Phys. Rev. D",
    volume = "87",
    number = "12",
    pages = "122003",
    year = "2013"
}

@article{Romano2016,
    author = "Romano, Joseph D. and Cornish, Neil J.",
    title = "{Detection methods for stochastic gravitational-wave backgrounds: a unified treatment}",
    eprint = "1608.06889",
    archivePrefix = "arXiv",
    primaryClass = "gr-qc",
    doi = "10.1007/s41114-017-0004-1",
    journal = "Living Rev. Rel.",
    volume = "20",
    number = "1",
    pages = "2",
    year = "2017"
}

@article{Cornish2025,
    author = "Cornish, Neil J.",
    title = "{Non-stationary noise in gravitational wave analyses: The wavelet domain noise covariance matrix}",
    eprint = "2511.10632",
    archivePrefix = "arXiv",
    primaryClass = "gr-qc",
    month = "11",
    year = "2025"
}

@article{Cook2006,
  title={Validation of Software for Bayesian Models Using Posterior Quantiles},
  author={Samantha R. Cook and Andrew Gelman and Donald B. Rubin},
  journal={Journal of Computational and Graphical Statistics},
  year={2006},
  volume={15},
  pages={675 - 692},
  url={https://api.semanticscholar.org/CorpusID:18161454}
}

@article{Gair2022,
    author = "Gair, Jonathan R. and others",
    title = "{The Hitchhiker{\textquoteright}s Guide to the Galaxy Catalog Approach for Dark Siren Gravitational-wave Cosmology}",
    eprint = "2212.08694",
    archivePrefix = "arXiv",
    primaryClass = "gr-qc",
    doi = "10.3847/1538-3881/acca78",
    journal = "Astron. J.",
    volume = "166",
    number = "1",
    pages = "22",
    year = "2023"
}

@article{Wouters2024,
    author = "Wouters, Thibeau and Pang, Peter T. H. and Dietrich, Tim and Van Den Broeck, Chris",
    title = "{Robust parameter estimation within minutes on gravitational wave signals from binary neutron star inspirals}",
    eprint = "2404.11397",
    archivePrefix = "arXiv",
    primaryClass = "astro-ph.IM",
    doi = "10.1103/PhysRevD.110.083033",
    journal = "Phys. Rev. D",
    volume = "110",
    number = "8",
    pages = "083033",
    year = "2024"
}

@article{Gupta2023,
    author = "Gupta, Toral and Cornish, Neil J.",
    title = "{Bayesian power spectral estimation of gravitational wave detector noise revisited}",
    eprint = "2312.11808",
    archivePrefix = "arXiv",
    primaryClass = "gr-qc",
    doi = "10.1103/PhysRevD.109.064040",
    journal = "Phys. Rev. D",
    volume = "109",
    number = "6",
    pages = "064040",
    year = "2024"
}

\end{document}